\documentclass[
 amsmath,amssymb,
 aps,
prd,
]{revtex4-2}

\usepackage{graphicx}
\usepackage{dcolumn}
\usepackage{bm}
\usepackage{amsmath,amssymb,amsthm,bm,bigdelim,multirow}
\usepackage{color}
\usepackage{booktabs}
\usepackage{hyperref}
\usepackage{dsfont}
\usepackage{here}
\usepackage{subcaption}
\usepackage{mathtools}
\usepackage{siunitx}
\allowdisplaybreaks

\newcommand{\kslash}{k\kern-1ex /}
\newcommand{\pslash}{p\kern-1ex /}
\newcommand{\qslash}{q\kern-1ex /}
\newcommand{\lslash}{l\kern-1ex /}
\newcommand{\sslash}{s\kern-1ex /}
\newcommand{\Dslash}{D\kern-1.2ex /}

\newcommand{\beqa}{\begin{eqnarray}}
\newcommand{\eeqa}{\end{eqnarray}}

\newcommand{\bd}{\begin{description}}
\newcommand{\ed}{\end{description}}

\newcommand{\ben}{\begin{eqnarray}}
\newcommand{\een}{\end{eqnarray}}
\newcommand{\nn}{\nonumber}
\def\lsim{\raise0.3ex\hbox{$<$\kern-0.75em\raise-1.1ex\hbox{$\sim$}}}
\def\gsim{\raise0.3ex\hbox{$>$\kern-0.75em\raise-1.1ex\hbox{$\sim$}}}
\def\simgt{\rlap{\lower 3.5 pt\hbox{$\mathchar \sim$}}\raise 2.0pt \hbox {$>$}}
\def\simlt{\rlap{\lower 3.5 pt\hbox{$\mathchar \sim$}}\raise 2.0pt \hbox {$<$}}

\usepackage{tikz,xcolor}
\definecolor{lime}{HTML}{A6CE39}
\DeclareRobustCommand{\orcidicon}{%
	\begin{tikzpicture}
	\draw[lime, fill=lime] (0,0) 
	circle [radius=0.16] 
	node[white] {{\fontfamily{qag}\selectfont \tiny ID}};	\draw[white, fill=white] (-0.0625,0.095) 
	circle [radius=0.007];	\end{tikzpicture}
	\hspace{-2mm}}
\foreach \x in {A,...,Z}{%
	\expandafter\xdef\csname orcid\x\endcsname{\noexpand\href{https://orcid.org/\csname orcidauthor\x\endcsname}{\noexpand\orcidicon}}
}

\begin{document}

\preprint{APS/123-QED}

\title{Phase structure of (3+1)-dimensional dense two-color QCD at $T=0$ in the strong coupling limit with the tensor renormalization group\\}
\author{Yuto Sugimoto\orcidA{}\,}
\email{sugimoto@nucl.phys.tohoku.ac.jp}
\affiliation{Department of Physics, Tohoku University, Sendai 980-8578, Japan}

\author{Shinichiro Akiyama\orcidB{}\,}
\email{akiyama@ccs.tsukuba.ac.jp}
\affiliation{Center for Computational Sciences, University of Tsukuba, Tsukuba, Ibaraki 305-8577, Japan}
\affiliation{Graduate School of Science, The University of Tokyo, Bunkyo-ku, Tokyo, 113-0033, Japan}

\author{Yoshinobu Kuramashi\orcidC{}\,}
\email{kuramasi@het.ph.tsukuba.ac.jp}
\affiliation{Center for Computational Sciences, University of Tsukuba, Tsukuba, Ibaraki 305-8577, Japan}
\date{\today}

\begin{abstract}
We investigate the phase structure of the (3+1)-dimensional strong coupling two-color QCD at zero temperature ($T=0$) with finite chemical potential using the tensor renormalization group method. The chiral and diquark condensates and the quark number density are evaluated as a function of the chemical potential. We further determine the critical exponents associated with the diquark condensate, which suggest consistency with the predictions of mean-field theory.
\end{abstract}
\date{\today}

\preprint{UTHEP-811, UTCCS-P-169}

\maketitle


\section{Introduction}
\label{sec:intro}

The complex action problem with finite chemical potential $\mu$ does not allow the standard Monte Carlo simulation to investigate the phase structure of QCD in the finite baryon density at $T=0$. On the other hand, the simulation of two-color QCD (QC$_2$D) with $\mu \ne 0$ is free from the complex action problem, so that it provides us an opportunity to understand the physics of matter under the finite density condition. 
In the strong coupling limit, the analytical methods have been used to investigate the phase structure.
In the 1980s, there were several mean field (MF) studies on chiral and diquark condensates in the strong coupling limit of the finite density QC$_2$D with the Kogut-Susskind quark~\cite{Dagotto:1986gw,Dagotto:1986xt,Klatke:1989xy}.  The results were later refined by Nishida, including the QCD case~\cite{Nishida:2003uj,Nishida:2003fb}. 
On the other hand, at finite coupling the standard Monte Carlo approaches have been employed since the mid-1980s~\cite{Nakamura:1984uz}. Although large-scale simulations are attainable in recent years, systematic studies changing the lattice sizes reveal temperature dependence of the phase structure~\cite{Boz:2019enj,Begun:2022bxj,Iida:2024irv}.
It is manifest that larger lattice sizes are required toward zero temperature limit.

In this paper we investigate the phase structure in the strong coupling limit of the (3+1)-dimensional ((3+1)$d$) finite density QC$_2$D at $T=0$ using the tensor renormalization group (TRG) method~\footnote{In this paper, the ``TRG method'' or the ``TRG approach'' refers to not only the original numerical algorithm proposed by Levin and Nave \cite{Levin:2006jai} but also its extensions~\cite{PhysRevB.86.045139,Shimizu:2014uva,PhysRevLett.115.180405,Sakai:2017jwp,PhysRevLett.118.110504,Hauru:2017tne,Adachi:2019paf,Kadoh:2019kqk,Akiyama:2020soe,PhysRevB.105.L060402,Akiyama:2022pse}.}. With the use of the Kogut-Susskind quark, we measure the chiral and diquark condensates and the quark number density as a function of $\mu$. 
There are two purposes in this work. First, we enlarge the lattice size up to $1024^4$ with the TRG method to investigate the phase structure at $T=0$, where the standard Monte Carlo approach is hardly accessible. It is useful to compare our results with the analytical ones obtained by the MF method and the $1/d_s$ expansion with $d_s$ the spatial dimension~\cite{Nishida:2003uj}.
Second, this is an appropriate preparatory study before exploring (3+1)$d$ finite density QCD, whose complex action problem prohibits the standard Monte Carlo approaches. The TRG method
may be qualified to investigate the phase structure of finite density QCD: This method, which is in principle free from the sign problem or the complex action problem, has been successfully applied to various types of models with these problems~\cite{Shimizu:2014uva,Shimizu:2014fsa,Kawauchi:2016xng,Kawauchi:2017dnj,Yang:2015rra,Shimizu:2017onf,Takeda:2014vwa,Kadoh:2018hqq,Kadoh:2019ube,Takeda:2019idb,Kuramashi:2019cgs,Akiyama:2020ntf,Akiyama:2020soe,Nakayama:2021iyp,Bloch:2021mjw,Bloch:2022vqz,Akiyama:2023hvt,Akiyama:2024qer,Luo:2022eje,Hite:2024ulb,Luo:2024lbh,Luo:2025qtv}. 
So far there exist several TRG applications, particularly dealing with the (1+1)$d$ QCD-like theories~\cite{Asaduzzaman:2023pyz,Pai:2024tip}, the (1+1)$d$ QCD in the strong coupling limit~\cite{Bloch:2022vqz}, and an effective field theory of the QCD at finite density~\cite{Akiyama:2020soe}.
In Ref.~\cite{Pai:2024tip}, the bond-weighted TRG (BTRG) algorithm~\cite{PhysRevB.105.L060402,Akiyama:2022pse} is employed to study the finite-density QC$_{2}$D. 
Although the tensor network formulation of the (3+1)$d$ QC$_2$D can be straightforwardly obtained by following Ref.~\cite{Pai:2024tip}, we need to resort a different TRG algorithm rather than the BTRG, which assumes the (1+1)$d$ tensor network representation.
A practically useful algorithm for the (3+1)$d$ QC$_2$D can be the anisotropic TRG algorithm~\cite{Adachi:2019paf,Oba:2019csk}, which has been already applied to the (3+1)$d$ finite density Nambu$-$Jona-Lasinio (NJL) model~\cite{Akiyama:2020soe}.
Although computational techniques treating four-fermi interactions are in common between the NJL model and the strong coupling QC$_2$D, the introduction of the color degree of freedom significantly increases the computational difficulty enlarging the size of the initial tensor by $2^{16}$.  Since this difficulty is further enhanced drastically from QC$_2$D to QCD we need to develop an efficient method to treat the large size of the initial tensor. 

This paper is organized as follows. In Sec.~\ref{sec:method}, we define the action of the QC$_2$D with finite chemical potential and give the tensor network representation. 
In Sec.~\ref{sec:results} we measure the chiral and diquark condensates and the quark number density as a function of the chemical potential. The critical exponents are also determined. The results are compared with the analytical ones in Ref.~\cite{Nishida:2003uj}.
Section~\ref{sec:summary} is devoted to the summary and outlook.

\section{Formulation and numerical algorithm}
\label{sec:method}
\subsection{Strong coupling QC$_2$D}
We consider the finite density QC$_2$D on a (3+1)$d$ lattice $\Lambda_{3+1}=\{(n_1,n_2,n_3,n_4)\ \vert n_{1,2,3}=1,\dots,N_s\, , n_4=1,\dots,N_\tau\}$ whose volume is $V=N_s^3\times N_\tau$ and temperature is defined as $T=1/N_\tau$. The lattice spacing $a$ is set to $a=1$ unless necessary. The action consists of the gluonic part and the fermionic one. For the former, the simple plaquette action is given by
\begin{equation}
  S_G= \frac{2 N_c}{g^2}\sum_{n,\nu,\rho}\left\{1-\frac{1}{N_c}{\rm Re Tr}\left(U_\nu(n)U_\rho(n+{\hat \nu})U_\nu^\dagger(n+{\hat \rho})U_\rho^\dagger(n)\right)\right\}
  \label{eq:action_g}
\end{equation}
with $N_c=2$ and $g$ the gauge coupling constant.   
The SU($N_{c}$)-valued link variable between the sites $n$ and $n+\hat{\nu}$, with $\hat{\nu}$ the unit vector in the $\nu$ direction, is denoted by $U_\nu(n)$. 
For the latter we employ the Kogut-Susskind quark action with the finite chemical potential $\mu$:
\begin{equation}
  S_F=\sum_n\left[\frac{1}{2}\sum_\nu \eta_\nu(n)\left\{e^{\mu\delta_{\nu,4}}{\bar \chi(n)}U_\nu(n)\chi(n+{\hat \nu})- e^{-\mu\delta_{\nu,4}}{\bar \chi(n+{\hat \nu})}U_\nu^\dagger(n)\chi(n)\right\}+m{\bar \chi(n)}\chi(n)\right]
  \label{eq:action_f}
\end{equation}
with $m$ the quark mass. $\chi=(\chi^1,\chi^2,\dots,\chi^{N_c})$ is a $N_c$ component Grassmann variable, and $\eta_\nu(n)$ is the staggered sign function defined by $\eta_\nu(n)=(-1)^{n_1+\cdots +n_{\nu-1}}$ with $\eta_1(n)=1$.

It is worthwhile to describe the symmetries in this theory.
Let us start with the $m=\mu=0$ case. Equation~(\ref{eq:action_f}), which corresponds to four-flavor quarks in the continuum limit, is invariant under the following continuous vector and axial vector transformation:
\ben
&&\chi(n) \rightarrow  e^{i\alpha}\chi(n),\;\;
\bar{\chi}(n) \rightarrow  \bar{\chi}(n)e^{-i\alpha}\\
&&\chi(n) \rightarrow  e^{i\alpha\epsilon(n)}\chi(n),\;\;
\bar{\chi}(n) \rightarrow  \bar{\chi}(n)e^{i\alpha\epsilon(n)}
\een
with $\alpha \in \mathbb{R}$ and $\epsilon(n)=(-1)^{n_1+n_2+n_3+n_4}$. The theory has the global U(1)$_{\rm V}\times$U(1)$_{\rm A}$ symmetry, which indicates the conservation of the baryon number and the axial charge. The two-color case is special in the sense that this symmetry is enlarged to the U(2) symmetry~\cite{Hands:1999md} thanks to the Pauli-G{\"u}rsey fermion-antifermion symmetry brought by the pseudoreality of the SU(2) group~\cite{Pauli:1957voo,Gursey:1958fzy}. The introduction of the finite chemical potential $\mu\ne 0$ at $m=0$ reduces the U(2) symmetry to the U(1)$_{\rm V}\times$U(1)$_{\rm A}$ symmetry. In the case of $m\ne 0$ and $\mu\ne 0$ we are left with the U(1)$_{\rm V}$ symmetry.
 
The partition function is defined as
\begin{equation}\label{eq:partition_function}
  Z=\int \left(\prod_{n\in\Lambda_{3+1},\nu}d\chi(n)d\bar{\chi}(n)dU_\nu(n)\right)e^{-S_G-S_F-S_D},
\end{equation}
where we add an explicit U(1)$_{\rm V}$ symmetry breaking term $S_D$ to detect the diquark condensation~\cite{Hands:1999md,Kogut:2001na}:
\begin{equation}
S_D=-\frac{\lambda}{2}\{\chi^t \tau_2 \chi+\bar{\chi} \tau_2 {\bar{\chi}}^t\} 
\end{equation}
with $\tau_2$ the second Pauli matrix acting on the color space. 
In the strong coupling limit $g\rightarrow \infty$, the gluonic action vanishes because of the overall factor of $1/g^2$, and the link variables are left only in the fermionic action. 

\subsection{Grassman tensor network representation}
In order to apply tensor network methods to this system, 
it is necessary to rewrite the partition function $Z$ in Eq.~\eqref{eq:partition_function} as a product of local tensors. 
Since the theory involves fermionic degrees of freedom, we must properly incorporate their anticommuting nature. 
To this end, we employ a Grassmann tensor network formulation~\cite{Akiyama:2020sfo}, 
in which the local tensors are defined using auxiliary Grassmann variables.  

For clarity, let us consider the case $\lambda=0$, which simplifies the discussion. 
The extension to $\lambda \neq 0$ is straightforward. 
To obtain the Grassmann tensor network representation of the partition function in Eq.~(\ref{eq:partition_function}), 
we introduce two-component auxiliary Grassmann variables 
$\zeta_\nu$ and $\xi_\nu$ $(\nu=1,2,3,4)$~\cite{Akiyama:2020sfo}. 
As a basic tool, for any $N$-component Grassmann variable $\theta_i$, one can use the identity
\begin{equation}
   \int_{\bar{\theta},\theta}\coloneqq \int \prod_{i=1}^{N} d\bar{\theta}_i\, d\theta_i e^{\theta_i \bar{\theta}_i}=1.
  \label{eq:grassmann_identity}
\end{equation}
By inserting Eq.~\eqref{eq:grassmann_identity} into the hopping terms of $S_F$ for each of the auxiliary variables $\zeta_\nu$ and $\xi_\nu$,
\begin{align}
&\exp\left[-\frac{\eta_\nu(n)}{2}e^{\mu\delta_{\nu,4}}\bar{\chi}(n)U_\nu(n){\chi}(n+\hat{\nu})\right] \nonumber\\
&= \iint_{\bar{\zeta}_\nu(n),\zeta_\nu(n)}\exp\bigg[-\frac{\eta_\nu(n)}{2}e^{\mu\delta_{\nu,4}}\bar{\chi}(n)\zeta_\nu(n)+\bar{\zeta}_\nu(n)U_\nu(n)\chi(n+\hat{\nu})\bigg]
\end{align}
\begin{align}
&\exp\left[\frac{\eta_\nu(n)}{2}e^{-\mu\delta_{\nu,4}}\bar{\chi}(n+\hat{\nu})U^\dagger_\nu(n)\chi(n)\right] \nonumber\\
&= \iint_{\bar{\xi}_\nu(n),\xi_\nu(n)}\exp\left[\bar{\chi}(n+\hat{\nu})U^\dagger_\nu(n)\bar{\xi}_\nu(n)-\frac{\eta_\nu(n)}{2}e^{-\mu\delta_{\nu,4}}\xi_\nu(n)\chi(n)\right],
\end{align}
we can define the local Grassmann tensor $\mathcal{T}$ by integrating out link variables $U_\nu(n)$ and the staggered fields $\chi(n),\bar{\chi}(n)$ at each lattice site $n$,
\begin{align}
\mathcal{T}&=\int \left[\prod_\nu dU_\nu\right] d\chi d\bar{\chi} e^{-m\bar{\chi}\chi}\prod_\nu\Biggl\{\exp\bigg[-\frac{\eta_\nu(n)}{2}e^{\mu\delta_{\nu,4}}\bar{\chi}\zeta_\nu(n)+\bar{\zeta}_\nu(n-\hat{\nu})U_\nu(n-\hat{\nu})\chi\bigg]\nonumber\\
&\hspace{150pt}\times\exp\left[\bar{\chi}U^\dagger_\nu(n-\hat{\nu})\bar{\xi}_\nu(n-\hat{\nu})-\frac{\eta_\nu(n)}{2}e^{-\mu\delta_{\nu,4}}\xi_\nu(n)\chi\right]\Biggr\}.\label{eq:local_tensor}\\
&=\sum_{\{i,j\}}\int \left[\prod_\nu dU_\nu\right] d\chi^1 d\bar{\chi}^1 d\chi^2 d\bar{\chi}^2 e^{-m\bar{\chi}\chi}\left[\prod_\nu\left(-\frac{\eta_\nu(n)}{2}e^{\mu\delta_{\nu,4}} \right)^{i^1_\nu+i^{2}_\nu}\left(-\frac{\eta_\nu(n)}{2}e^{-\mu\delta_{\nu,4}} \right)^{j^1_\nu+j^{2}_\nu}\right]\nonumber\\
&\hspace{20pt}\times\left[\prod_\nu \left(\sum_{a_\nu}\bar{\zeta}^2_\nu U^{2a_\nu}_\nu \chi^{a_\nu}\right)^{i'^2_\nu}\right]\left[\prod_\nu\left(\sum_{b_\nu}\bar{\zeta}^1_\nu U^{1b_\nu}_\nu \chi^{b_\nu}\right)^{i'^1_\nu}\right]\left[\prod_\nu\left(\xi^1_\nu\chi^1\right)^{j^1_\nu}\right]\left[\prod_\nu\left(\xi^2_\nu\chi^2\right)^{j^2_\nu}\right]\nonumber\\
&\hspace{20pt}\times\left[\prod_\nu\left(\sum_{c_\nu}\bar{\chi}^{c_\nu} U^{c_\nu2\dagger}_\nu\bar{\xi}^2_\nu  \right)^{j'^2_\nu}\right]\left[\prod_\nu\left(\sum_{d_\nu}\bar{\chi}^{d_\nu} U^{d_\nu1\dagger}_\nu\bar{\xi}^1_\nu  \right)^{j'^1_\nu}\right]\left[\prod_\nu\left(\bar{\chi}^1\zeta^1_\nu\right)^{i^1_\nu}\right]\left[\prod_\nu\left(\bar{\chi}^2\zeta^2_\nu\right)^{i^2_\nu}\right]\label{eq:local_tensor2}.
\end{align}
In the above, we have introduced the fermion occupation numbers 
$i'^{1}_{\nu}, i'^{2}_{\nu}, j'^{2}_{\nu}, j'^{1}_{\nu}, 
 j^{1}_{\nu}, j^{2}_{\nu}, i^{1}_{\nu}, i^{2}_{\nu} \in \{0,1\}$ 
$(\nu=1,2,3,4)$, which arise from the Taylor expansion of 
Eq.~(\ref{eq:local_tensor}), and the auxiliary integer variables 
$a_{\nu}, b_{\nu}, c_{\nu}, d_{\nu} \in \{1,2\}$ to sum out the color 
indices of the staggered fields $\chi$ connected to the link variables 
$U_{\nu}$. The summation $\sum_{\{i,j\}}$ denotes 
$\left(\prod_\nu \sum_{i'^{1}_{\nu}, i'^{2}_{\nu}, j'^{1}_{\nu}, j'^{2}_{\nu}, 
 j^{1}_{\nu}, j^{2}_{\nu}, i^{1}_{\nu}, i^{2}_{\nu}}\right)$. 
Here we omit the explicit spacetime dependence except for $\eta_\nu(n)$. 
This implies that there are eight types of initial tensors, 
depending on the staggered sign~\cite{Akiyama:2020soe}.

To carry out the group integral over $U_\nu$ in Eq.~(\ref{eq:local_tensor2}), we introduce the indicator function 
$g(i) = \tfrac{1}{2}\delta_{i,0} + \delta_{i,1}$. 
Using this function, the Grassmann variables and the link variables can be 
separated as follows:
\begin{align}
    \left(\sum_{a_\nu}\bar{\zeta}^2_\nu U^{2a_\nu}_\nu \chi^{a_\nu}\right)^{i'^2_\nu}
    &= g(i'^2_\nu) \sum_{a_\nu} 
       \left[(\bar{\zeta}^2_\nu)^{i'^2_\nu}\,
             (U^{2a_\nu}_\nu)^{i'^2_\nu}\,
             (\chi^{a_\nu})^{i'^2_\nu}\right].
\end{align}
Thus, within the sum of $\sum_{a_\nu,b_\nu,c_\nu,d_\nu}$, we can analytically integrate the link variables,
\begin{equation}
       F^{a_\nu b_\nu c_\nu d_\nu}_{i'^2_\nu i'^1_\nu j'^2_\nu j'^1_\nu}=g(i'^2_\nu)g(i'^1_\nu)g(j'^2_\nu)g(j'^1_\nu)\int dU_\nu(U^{2a_\nu}_\nu)^{i'^2_\nu}(U^{1b_\nu}_\nu)^{i'^1_\nu}(U^{c_\nu2\dagger}_\nu)^{j'^2_\nu}(U^{d_\nu1\dagger}_\nu)^{j'^1_\nu},\label{eq:F_abcd}
\end{equation}
where $F^{a_\nu b_\nu c_\nu d_\nu}_{i'^2_\nu i'^1_\nu j'^2_\nu j'^1_\nu}$ indicates the results of the integration using Weingarten calculus~\cite{Weingarten:1977ya}.
After integrating out the gauge fields, all that remains is to integrate out $\chi$ and $\bar{\chi}$. 
A straightforward calculation yields the final form of the Grassmann tensor $\mathcal{T}$:
\begin{align}
    \hspace*{-2cm}&\mathcal{T}_{\Psi_1 (n)\Psi_2(n)\Psi_3(n)\Psi_4(n)\bar{\Psi}_4(n-\hat{4})\bar{\Psi}_3(n-\hat{2})\bar{\Psi}_2(n-\hat{2})\bar{\Psi}_1 (n-\hat{1})}\nonumber\\    &=\sum_{i,j,i',j'}T_{\left(i^1_1,i^2_1,j^1_1,j^2_1\right)\left(i^1_2,i^2_2,j^1_2,j^2_2\right)\left(i^1_3,i^2_3,j^1_3,j^2_3\right)\left(i^1_4,i^2_4,j^1_4,j^2_4\right)\left(i'^1_1,i'^2_1,j'^1_1,j'^2_1\right)\left(i'^1_2,i'^2_2,j'^1_2,j'^2_2\right)\left(i'^1_3,i'^2_3,j'^1_3,j'^2_3\right)\left(i'^1_4,i'^2_4,j'^1_4,j'^2_4\right)}\nonumber\\
    \times&\left[(\zeta^1_1)^{i^1_1}(\zeta^2_1)^{i^2_1}(\xi^1_1)^{j^1_1}(\xi^2_1)^{j^2_1}\right]\left[(\zeta^1_2)^{i^1_2}(\zeta^2_2)^{i^2_2}(\xi^1_2)^{j^1_2}(\xi^2_2)^{j^2_2}\right]\left[(\zeta^1_3)^{i^1_3}(\zeta^2_3)^{i^2_3}(\xi^1_3)^{j^1_3}(\xi^2_3)^{j^2_3}\right]\left[(\zeta^1_4)^{i^1_4}(\zeta^2_4)^{i^2_4}(\xi^1_4)^{j^1_4}(\xi^2_4)^{j^2_4}\right]\nonumber\\
    &\hspace*{-1cm}\times\left[(\bar{\xi}^2_4)^{j'^2_4}(\bar{\xi}^1_4)^{j'^1_4}(\bar{\zeta}^2_4)^{i'^2_4}(\bar{\zeta}^1_4)^{i'^1_4}\right]\left[(\bar{\xi}^2_3)^{j'^2_3}(\bar{\xi}^1_3)^{j'^1_3}(\bar{\zeta}^2_3)^{i'^2_3}(\bar{\zeta}^1_3)^{i'^1_3}\right]\left[(\bar{\xi}^2_2)^{j'^2_2}(\bar{\xi}^1_2)^{j'^1_2}(\bar{\zeta}^2_2)^{i'^2_2}(\bar{\zeta}^1_2)^{i'^1_2}\right]\left[(\bar{\xi}^2_1)^{j'^2_1}(\bar{\xi}^1_1)^{j'^1_1}(\bar{\zeta}^2_1)^{i'^2_1}(\bar{\zeta}^1_1)^{i'^1_1}\right],\label{eq:GrassmanT}
\end{align}
where $\Psi_\nu=(\zeta^1_\nu,\zeta^2_\nu,\xi^1_\nu,\xi^2_\nu)$ and 
$\bar{\Psi}_\nu=(\bar{\zeta}^1_\nu,\bar{\zeta}^2_\nu,\bar{\xi}^1_\nu,\bar{\xi}^2_\nu)$ denote the composite Grassmann variables, and the coefficient tensor $T$ is defined as
\begin{align}\label{eq:coef}
&T_{\left(i^1_1,i^2_1,j^1_1,j^2_1\right)\left(i^1_2,i^2_2,j^1_2,j^2_2\right)\left(i^1_3,i^2_3,j^1_3,j^2_3\right)\left(i^1_4,i^2_4,j^1_4,j^2_4\right)\left(i'^1_1,i'^2_1,j'^1_1,j'^2_1\right)\left(i'^1_2,i'^2_2,j'^1_2,j'^2_2\right)\left(i'^1_3,i'^2_3,j'^1_3,j'^2_3\right)\left(i'^1_4,i'^2_4,j'^1_4,j'^2_4\right)}\nonumber\\
&=\sum_{i,j}\left[\prod_\nu\left(-\frac{\eta_\nu(n)}{2}e^{\mu\delta_{\nu,4}} \right)^{i^1_\nu+i^{2}_\nu}\left(-\frac{\eta_\nu(n)}{2}e^{-\mu\delta_{\nu,4}} \right)^{j^1_\nu+j^{2}_\nu}\right]\left(\prod_\nu \sum_{a_\nu,b_\nu,c_\nu,d_\nu}\right)\left(\prod_\nu \tilde{F}^{a_\nu b_\nu c_\nu d_\nu}_{i'^2_\nu i'^1_\nu j'^2_\nu j'^1_\nu}\right)\nonumber\\
&\times \Big(-\delta_{1, \sum_\nu (\delta^{1,a_\nu}i'^2_\nu+\delta^{1,b_\nu}i'^1_\nu+j^1_\nu)}\delta_{1, \sum_\nu (\delta^{2,a_\nu}i'^2_\nu+\delta^{2,b_\nu}i'^1_\nu+j^2_\nu)}\delta_{1, \sum_\nu (\delta^{1,c_\nu}j'^2_\nu+\delta^{1,d_\nu}j'^1_\nu+i^1_\nu)}\delta_{1, \sum_\nu (\delta^{2,c_\nu}j'^2_\nu+\delta^{2,d_\nu}j'^1_\nu+i^2_\nu)}\nonumber\\
& +m\delta_{1, \sum_\nu (\delta^{1,a_\nu}i'^2_\nu+\delta^{1,b_\nu}i'^1_\nu+j^1_\nu)}\delta_{0, \sum_\nu (\delta^{2,a_\nu}i'^2_\nu+\delta^{2,b_\nu}i'^1_\nu+j^2_\nu)}\delta_{1, \sum_\nu (\delta^{1,c_\nu}j'^2_\nu+\delta^{1,d_\nu}j'^1_\nu+i^1_\nu)}\delta_{0, \sum_\nu (\delta^{2,c_\nu}j'^2_\nu+\delta^{2,d_\nu}j'^1_\nu+i^2_\nu)}\nonumber\\
&+m\delta_{0, \sum_\nu (\delta^{1,a_\nu}i'^2_\nu+\delta^{1,b_\nu}i'^1_\nu+j^1_\nu)}\delta_{1, \sum_\nu (\delta^{2,a_\nu}i'^2_\nu+\delta^{2,b_\nu}i'^1_\nu+j^2_\nu)}\delta_{0, \sum_\nu (\delta^{1,c_\nu}j'^2_\nu+\delta^{1,d_\nu}j'^1_\nu+i^1_\nu)}\delta_{1, \sum_\nu (\delta^{2,c_\nu}j'^2_\nu+\delta^{2,d_\nu}j'^1_\nu+i^2_\nu)}\nonumber\\
& +m^2 \delta_{0, \sum_\nu (i'^2_\nu+i'^1_\nu+j^1_\nu+j^2_\nu+j'^2_\nu+j'^1_\nu+i^1_\nu+i^2_\nu)}\Big)\nonumber\\
&(-1)^{R_{\left(i^1_1,i^2_1,j^1_1,j^2_1\right)\left(i^1_2,i^2_2,j^1_2,j^2_2\right)\left(i^1_3,i^2_3,j^1_3,j^2_3\right)\left(i^1_4,i^2_4,j^1_4,j^2_4\right)\left(i'^1_1,i'^2_1,j'^1_1,j'^2_1\right)\left(i'^1_2,i'^2_2,j'^1_2,j'^2_2\right)\left(i'^1_3,i'^2_3,j'^1_3,j'^2_3\right)\left(i'^1_4,i'^2_4,j'^1_4,j'^2_4\right)}}.\nonumber \\
\end{align}
The tensor $\tilde{F}$ 
is a redefined version of $F$ 
that incorporates $a,b,c,d$ dependence of the sign factors 
arising from the reordering of $\chi^2,\chi^1,\bar{\chi}^2,\bar{\chi}^1$ 
in the integration procedure. 
The tensor $R$ likewise includes sign factors originating from the 
rearrangement of the auxiliary Grassmann fields 
$\zeta,\xi,\bar{\zeta},\bar{\xi}$ to the form of Eq.~(\ref{eq:GrassmanT}).
The explicit definitions of $F$, $\tilde{F}$, and $R$ are provided in the Appendix. 
Using these definitions, the partition function $Z$ can be expressed as a Grassmann tensor network~\cite{Kadoh:2018hqq},
\begin{equation}\label{eq:Z_grassmann}
    Z = \mathrm{gTr} \left[\prod_{n} 
    \mathcal{T}_{\Psi_1 (n)\Psi_2 (n)\Psi_3 (n)\Psi_4 (n)\,
    \bar{\Psi}_4 (n-\hat{4})\bar{\Psi}_3 (n-\hat{3})\bar{\Psi}_2 (n-\hat{2})\bar{\Psi}_1 (n-\hat{1})
    }
    \right].
\end{equation}
Here, $\mathrm{gTr}$ denotes the Grassmann trace, i.e., the integration over the auxiliary Grassmann variables $\Psi$ and $\bar{\Psi}$ in the sense of Eq.~\eqref{eq:grassmann_identity}.

To evaluate Eq.~\eqref{eq:Z_grassmann}, we have employed the Grassmann-anisotropic TRG method~\cite{Adachi:2019paf,Akiyama:2020soe} with the bond dimension $D$ incorporating the multi-GPU parallelization technique proposed in Ref.~\cite{Sugimoto:2025xva}.
\section{Numerical results}
\label{sec:results}

\begin{figure}[htbp]
	\centering
	\includegraphics[width=0.7\hsize]{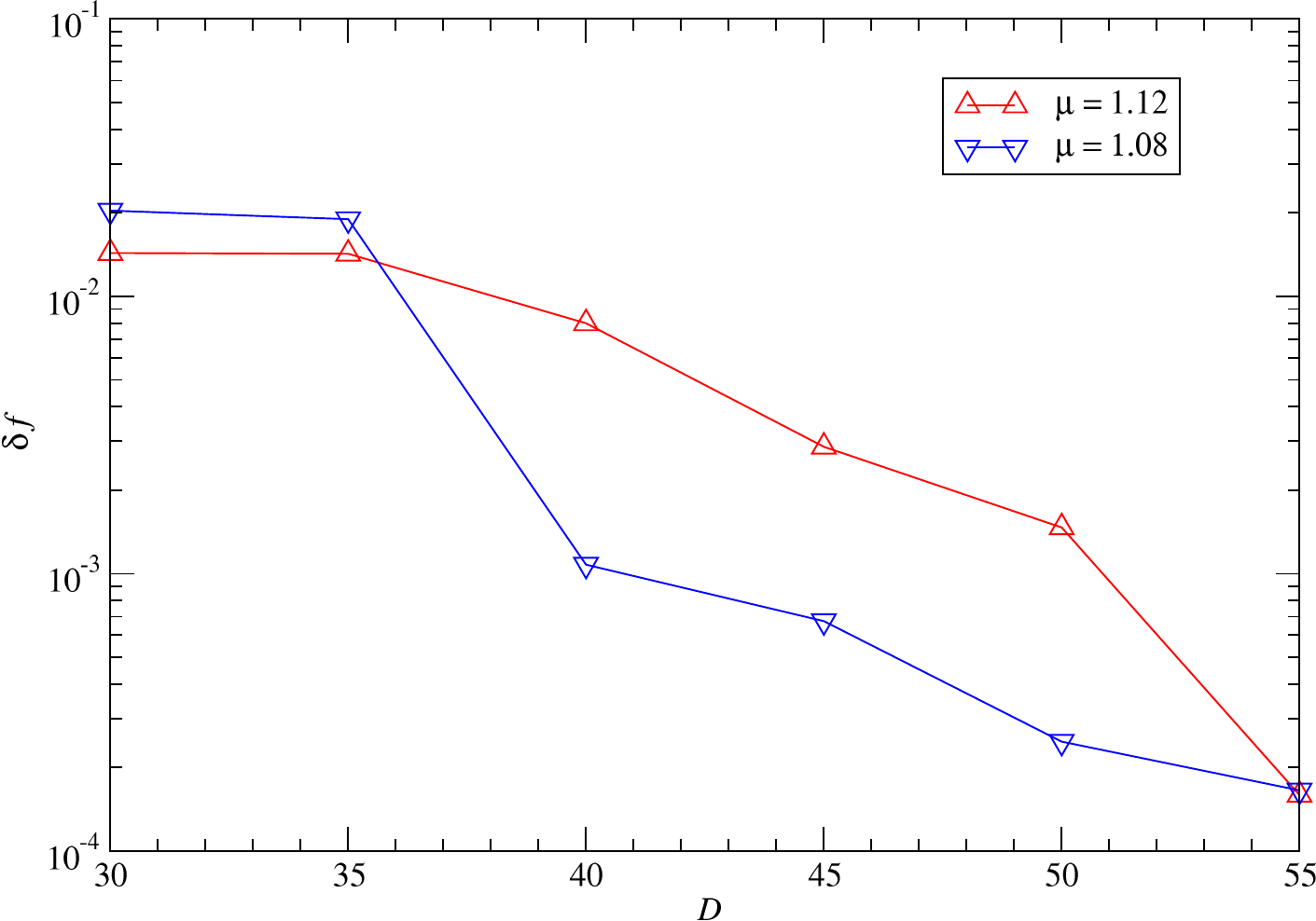}
	\caption{Convergence behaviors of thermodynamic potential as a function of $D$ at $(m,\lambda)=(1.0,0)$ on a $1024^4$.$\mu=1.08$ and 1.12 cases are plotted.}
  	\label{fig:delta_f}
\end{figure}

Before presenting the results for the physical quantities, we check the convergence behavior of the thermodynamic potential $f(m,\mu,\lambda,D)=\ln Z(m,\mu,\lambda,D)/V$ by defining the quantity
\begin{equation}
    \delta f(m,\mu,\lambda,D)=\left\vert \frac{\ln Z(m,\mu,\lambda,D)-\ln Z(m,\mu,\lambda,D=60)}{\ln Z(m,\mu,\lambda,D=60)}\right\vert
\end{equation}
 on $V=1024^4$. Figure~\ref{fig:delta_f} shows the $D$ dependence of $\delta f$ at $(m,\lambda)=(1.0,0)$ with the choices of $\mu=1.08$ and $1.12$, the former of which resides in the Silver Blaze region and the latter in the region of the diquark condensation, as shown below. We observe good convergence behaviors for both cases: $\delta f$ reaches $O(10^{-4})$ at $D=55$. Hereafter, we present the results at $D=55$ and omit the argument $D$ in the functions of $f$ and $Z$.

\subsection{Chiral condensate and number density}
\label{subsec:chiral}

We first show the results for the expectation values of the chiral condensate $\langle {\bar \chi} \chi\rangle$ and the quark number density $\langle n\rangle$ employing the numerical derivative of the thermodynamic potential:
\begin{align}
    \langle {\bar \chi} \chi \rangle 
    = \frac{1}{V}\frac{\partial \ln Z(m,\mu,\lambda=0)}{\partial m}
    \simeq\frac{1}{V}\frac{\ln Z(m+\Delta m,\mu,\lambda=0)-\ln Z(m-\Delta m,\mu,\lambda=0)}{2\Delta m},
\end{align}
\begin{align}
    \langle n \rangle 
    &= \frac{1}{V}\frac{\partial \ln Z(m,\mu,\lambda=0)}{\partial \mu}\nonumber\\
    &\simeq \frac{1}{V}\frac{\ln Z(m,\mu+\Delta \mu,\lambda=0)-\ln Z(m,\mu -\Delta \mu,\lambda=0)}{2\Delta \mu}
\end{align}
with $\Delta m=0.02$ and $\Delta \mu=0.02$. 

\begin{figure}[htbp]
	\centering
	\includegraphics[width=0.7\hsize]{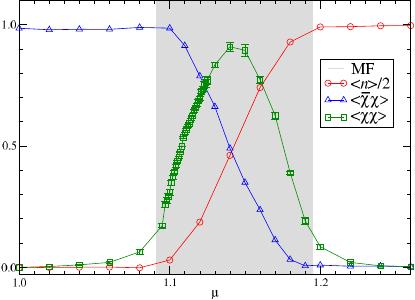}
	\caption{$\mu$ dependence of $\langle {\bar \chi} \chi\rangle$, $\langle \chi \chi\rangle$ and $\langle n\rangle/2$ at $(m,\lambda)=(1.0,0)$ on a $1024^4$ lattice with $D=55$.}
  	\label{fig:mu-dep}
\end{figure}

\begin{figure}[htbp]
	\centering
  \begin{minipage}{0.48\hsize}
    \centering
    \includegraphics[width=\hsize]{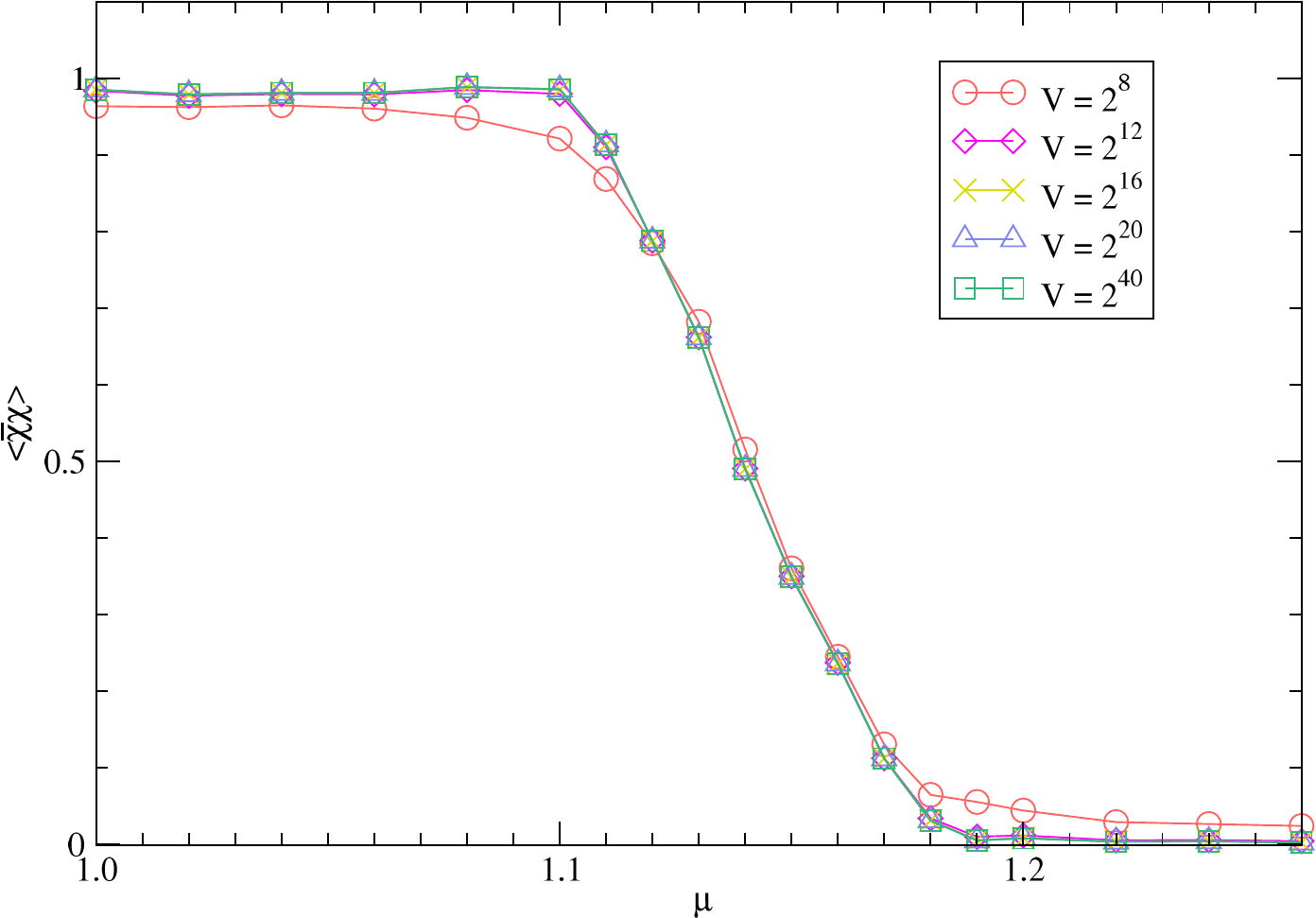}
  \end{minipage}
  \begin{minipage}{0.48\hsize}
    \centering
    \includegraphics[width=\hsize]{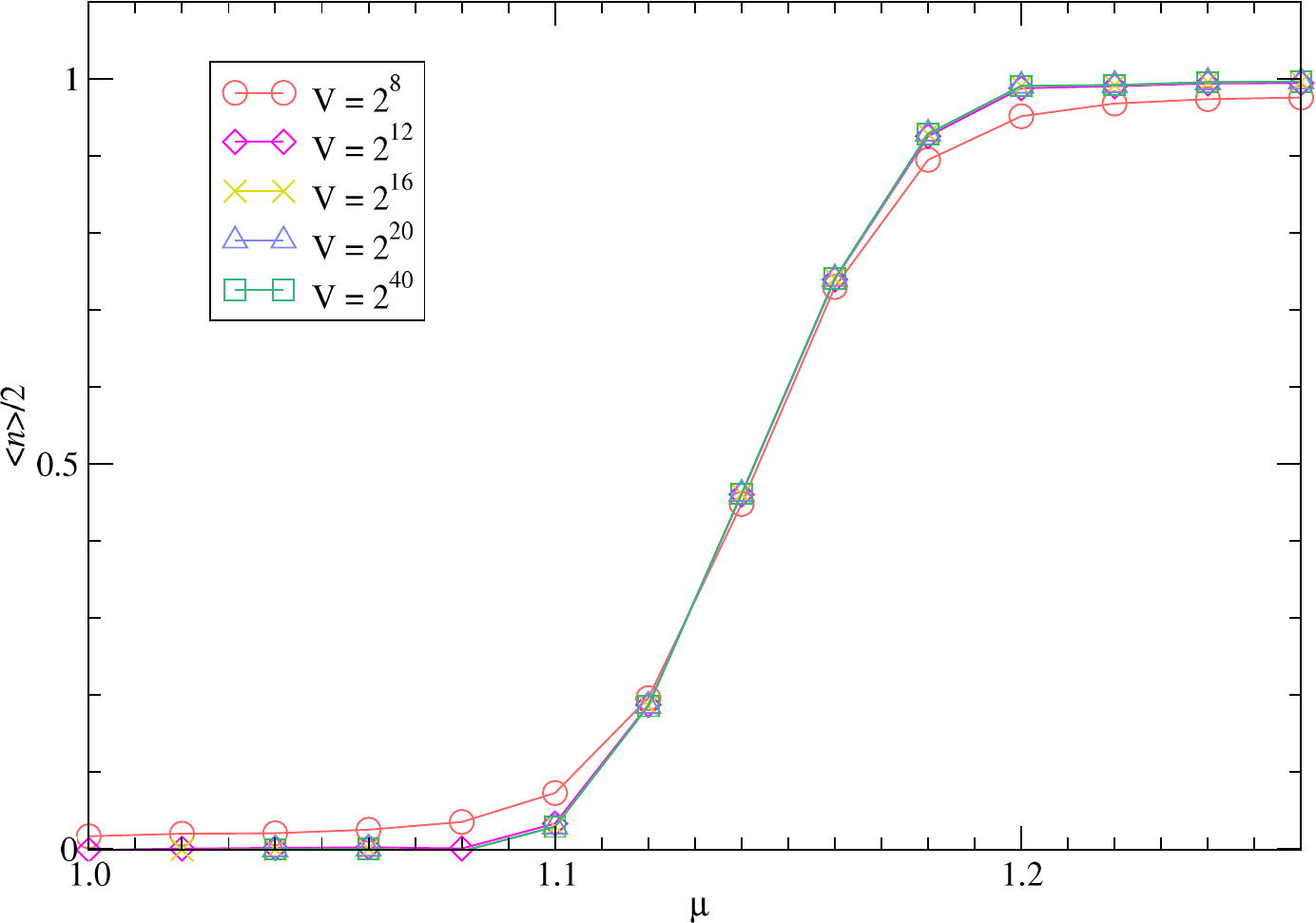}
  \end{minipage}
	\caption{Volume dependence of $\langle {\bar \chi} \chi\rangle$ and $\langle n\rangle/2$ at $(m,\lambda)=(1.0,0)$ with $D=55$.}
  	\label{fig:vol-dep_cn}
\end{figure}

Figure~\ref{fig:mu-dep} plots $\langle {\bar \chi} \chi\rangle$ and $\langle n\rangle$, together with the diquark condensate $\langle \chi\chi\rangle$, as a function of $\mu$ at $m=1.0$ on a $1024^4$ lattice with the choice of $D=55$. We find that the qualitative behaviors of $\langle {\bar \chi} \chi\rangle$ and $\langle n\rangle$  are similar to those for $m \ne 0$ given by the MF analysis in Ref.~\cite{Nishida:2003uj}: $\langle {\bar \chi} \chi\rangle$ stays constant up to $\mu=\mu_c^{\rm low}$ and monotonically decreases up to the vanishing point of $\mu=\mu_c^{\rm up}$, where $\mu_c^{\rm low} \le \mu \le \mu_c^{\rm up}$ is the region of $\langle\chi \chi\rangle \ne 0$; $\langle n\rangle/2$ starts to increase from zero at $\mu =\mu_c^{\rm low}$ and saturates to $\langle n\rangle/2=1$ at $\mu \ge \mu_c^{\rm up}$. Note that $\langle n\rangle>2$ is not allowed due to the Pauli exclusion principle.
In Fig.~\ref{fig:vol-dep_cn} we plot the results for $\langle {\bar \chi} \chi\rangle$ and $\langle n\rangle$ changing the volume. It is clear that the $1024^4$ lattice size is large enough to obtain the results in the thermodynamic limit at the zero temperature. 

\subsection{Diquark condensate}
\label{subsec:diquark}

\begin{figure}[htbp]
	\centering
	\includegraphics[width=0.7\hsize]{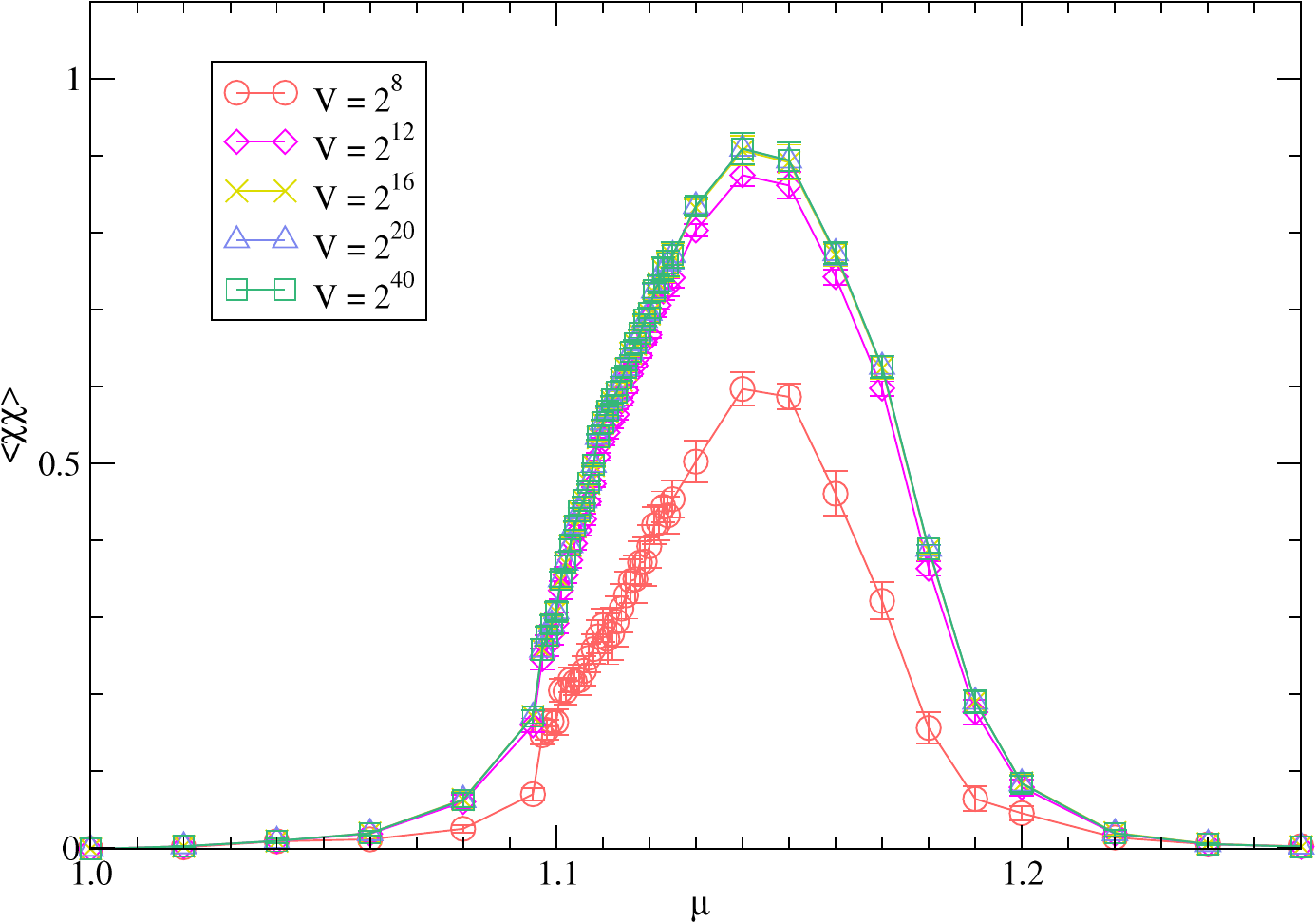}
	\caption{Volume dependence of $\langle \chi \chi\rangle$ at $(m,\lambda)=(1.0,0)$ with $D=55$.}
  	\label{fig:vol-dep_d}
\end{figure}

The diquark condensate is defined by
\begin{equation}
  \langle \chi\chi \rangle = \left.\frac{1}{V}\frac{\partial \ln Z(m,\mu,\lambda)}{\partial \lambda} \right\vert_{\lambda\rightarrow 0}. 
\end{equation}
In order to avoid instabilities of the numerical derivative in terms of $\lambda$ in the vicinity of $\lambda=0$, we evaluate the diquark condensate by fitting $f(m,\mu,\lambda)$ with the following function:
\begin{equation}
f(m,\mu,\lambda)=b_1\lambda^2+b_2|\lambda|+f(m,\mu,\lambda=0),   
\end{equation}
where the coefficient $b_2$ gives the diquark condensate. 
The range of $\lambda$ was chosen depending on $\mu$: 
for $\mu=1.095$, $\lambda\in [0, 0.030]$ with $\Delta \lambda=0.002$, for $1.097 \leq \mu \leq 1.11$, $\lambda\in [0, 0.045]$ with $\Delta\lambda=0.005$, and for other $\mu$, $\lambda\in [0, 0.040]$ with $\Delta\lambda=0.005$.
We find that the $\mu$ dependence of the diquark condensate in Fig.~\ref{fig:mu-dep} is qualitatively described by the MF analysis in Ref.~\cite{Nishida:2003uj}.
The region of $\langle \chi \chi\rangle \ne 0$ shows rough consistency with the MF prediction, which is denoted by the gray band from $\mu_c^{\rm low}=1.0913$ to $\mu_c^{\rm up}=1.1944$. 
Figure~\ref{fig:vol-dep_d} shows the volume dependence of the diquark condensate. We hardly observe the volume dependence for the results on $V\geq 2^{16}$. As in the case of $\langle {\bar \chi} \chi\rangle$ and $\langle n\rangle$, the diquark condensate in the thermodynamic limit at zero temperature is obtained on $V=1024^4$. 

\begin{figure}[htbp]
	\centering
	\includegraphics[width=0.7\hsize]{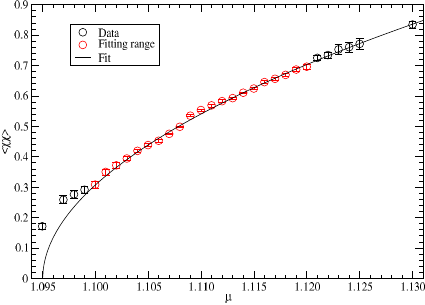}
	\caption{$\mu$ dependence of $\langle \chi \chi\rangle$ around  $\mu_c^{\rm low}$ at $(m,\lambda)=(1.0,0)$ on a $1024^4$ lattice with $D=55$.}
  	\label{fig:mu-fit}
\end{figure}

\begin{figure}[htbp]
	\centering
	\includegraphics[width=0.7\hsize]{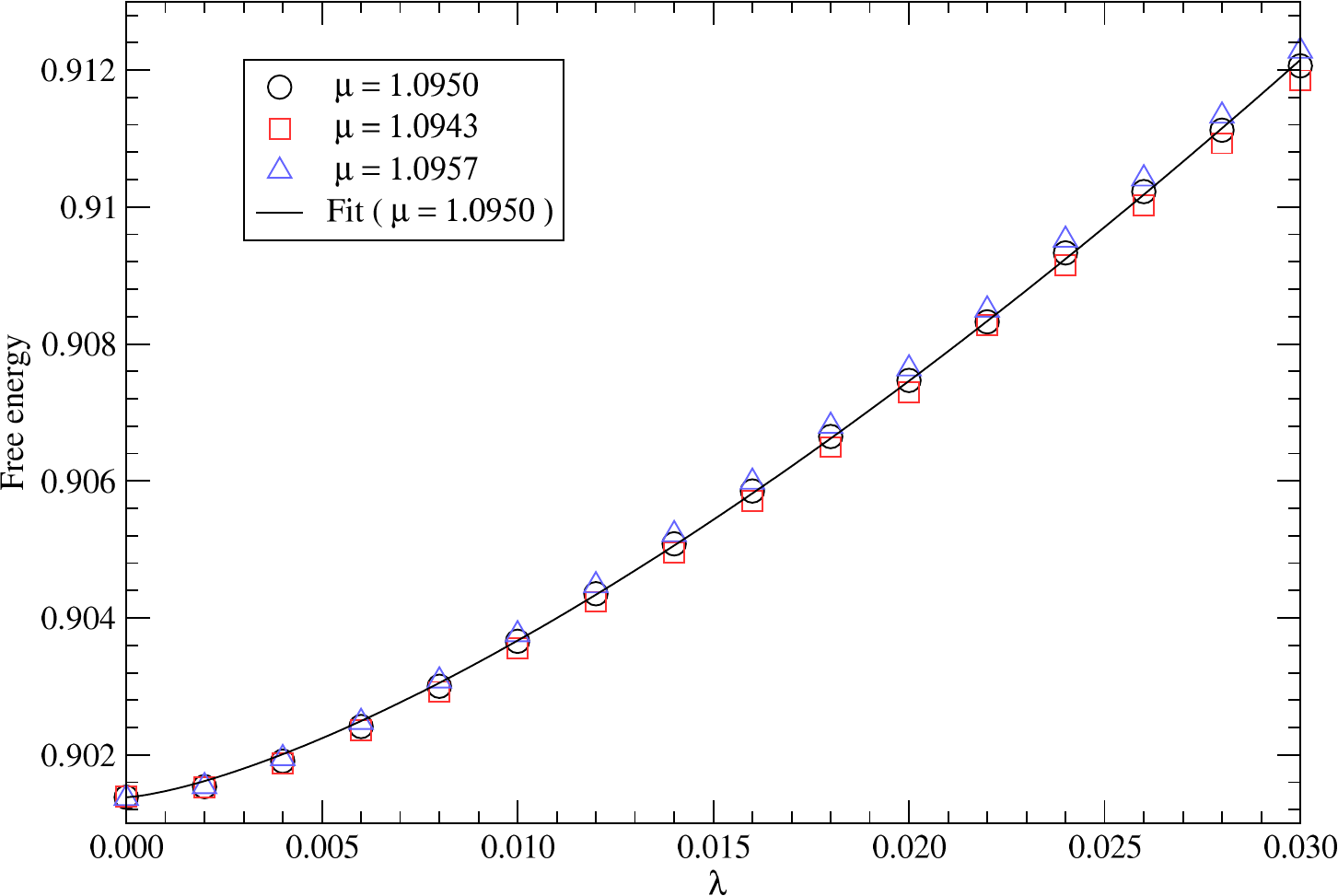}
	\caption{$\lambda$ dependence of $\langle \chi \chi\rangle$ at $m=1.0$ on a $1024^4$ lattice with $D=55$. Circle, triangle and square symbols denote the results at $\mu=1.0950$, 1.0957 and $1.0943$, respectively.}
  	\label{fig:lambda-fit}
\end{figure}

Now let us determine the critical exponents $\beta_m$ and $\delta$. Figure~\ref{fig:mu-fit} shows the $\mu$ dependence of $\langle \chi \chi\rangle$ with $D=55$ near the critical point $\mu_c^{\rm low}$. A fit of $\langle \chi \chi\rangle$ in the region of $\mu\in [1.10,1.12]$ assuming $\langle \chi \chi\rangle=A(\mu- \mu_c^{\rm low})^{\beta_m}$ with $A$, $\mu_c^{\rm low}$ and $\beta_m$ the free parameters yields $A=4.7(4)$, $\mu_c^{\rm low}=1.0950(7)$ and $\beta_m=0.514(27)$. The value of the critical $\mu$ is close to the MF prediction of $\mu_c^{\rm low}=1.0913$ and $\beta_m$ shows consistency with the MF prediction of $\beta_m=0.5$. The lower limit of the fitting range is chosen to avoid the immediate vicinity of the critical point. Below $\mu=1.10$ nonanalytic behaviors of $f(m,\mu,\lambda)$ in terms of $\lambda$ becomes manifest. The robustness of the fit results against the change of the upper limit of the fitting range is indicated by the fact that the data in the region of $1.12< \mu \le 1.13$ is on the fit curve. 
We also estimate another critical exponent $\delta$ from the $\lambda$ dependence of the free energy at $\mu_c^{\rm low}=1.0950(7)$ plotted in Fig.~\ref{fig:lambda-fit}. The circle, triangle and square symbols denote the results at $\mu=1.0950$, 1.0957 and 1.0943, respectively. The solid curve denotes the fit result at $\mu=1.0950$ with the form  $f(m,\mu,\lambda=0)+b_0\lambda^{1+1/\delta}$ choosing the fitting range of $\lambda\in [0,0.03]$. The free parameters are determined to be $(b_0,\delta)=(1.51(6),2.44(6))$ at $\mu=1.0950$, $(b_0,\delta)=(1.42(5),2.59(6))$ at $\mu=1.0957$ and  $(b_0,\delta)=(1.59(7),2.32(6))$ at $\mu=1.0943$. The scattering of the central value of $\delta$ due to the uncertainty in the determination of  $\mu_c^{\rm low}$ is considered to be the systematic error. The value of $\delta$ is estimated as $\delta=2.44(6)(^{+0.15}_{-0.12})$, where the numbers in the second parentheses represent the systematic error. The discrepancy between our result of $\delta$ and the MF prediction of $\delta=3$ may be attributed to the difficulty in the precise determination of $\mu_c^{\rm low}$. We find that the value of $\delta$ seems to become larger as $\mu$ increases in the vicinity of the critical point. Actually in case that $\delta$ is measured at $\mu=1.0971$, which is 3$\sigma$ away from $\mu_c=1.0950$, we obtain $\delta=2.91(6)$ resulting in almost consistent with the MF prediction.

\section{Summary and outlook} 
\label{sec:summary}

We have investigated the phase structure of the (3+1)$d$ finite density QC$_2$D at $m=1.0$ on a $1024^4$ lattice with $D=55$. 
The volume is large enough to be regarded as the thermodynamic limit at zero temperature. 
The TRG results for the chiral and diquark condensates and the number density show similar $\mu$ dependences to the previous MF results. 
The value of the critical exponents $\beta_m$ is comparable to the MF prediction, while $\delta$ shows a slightly smaller value.  

As a next step, it may be worthwhile to incorporate gauge fields at finite coupling. 
Another possible direction is an extension to the Wilson quark.

\begin{acknowledgments}
  Numerical calculation for the present work was carried out using the computational resources of SQUID provided by Osaka University through the HPCI System Research Project (Project No. hp250120). We also used Yukawa-21 at Yukawa Institute Computer Facility in Kyoto University and the supercomputer Pegasus under the Multidisciplinary Cooperative Research Program of Center for Computational Sciences, University of Tsukuba. 
  Y.\ S. acknowledges support from the Graduate Program on Physics for the Universe (GP-PU), Tohoku University, and from JSPS KAKENHI (Grant-in-Aid for JSPS Fellows) Grant No. 25KJ0537.
  S.\ A. acknowledges the support from JSPS KAKENHI Grant No. JP23K13096, and the Top Runners in Strategy of Transborder Advanced Researches (TRiSTAR) program conducted as the Strategic Professional Development Program for Young Researchers by the MEXT. 
  This work is supported in part by Grants-in-Aid for Scientific Research from the Ministry of Education, Culture, Sports, Science and Technology (MEXT) (Grants No. 24H00214, No. 24H00940).
\end{acknowledgments}

\appendix
\section{EXPLICIT FORM OF $F$ AND $R$}
We provide explicit form of the initial tensors.
The Weingarten integration of link variables is expressed as follows:
\begin{align}
F^{a_\nu b_\nu c_\nu d_\nu}_{i'^2_\nu i'^1_\nu j'^2_\nu j'^1_\nu}&=g(i'^2_\nu)g(i'^1_\nu)g(j'^2_\nu)g(j'^1_\nu)\int dU_\nu(U^{2a_\nu}_\nu)^{i'^2_\nu}(U^{1b_\nu}_\nu)^{i'^1_\nu}(U^{c_\nu2\dagger}_\nu)^{j'^2_\nu}(U^{d_\nu1\dagger}_\nu)^{j'^1_\nu}\nonumber\\
&=\bigg[ \delta_{0,i'^2_\nu}\delta_{0,i'^1_\nu}\delta_{0,j'^2_\nu} \delta_{0,j'^1_\nu}+\delta_{1,i'^2_\nu}\delta_{1,i'^1_\nu}\delta_{1,j'^2_\nu}\delta_{1,j'^1_\nu}\frac{1}{3}\left\{\delta^{2,2}\delta^{a_\nu ,c_\nu}\delta^{1,1}\delta^{b_\nu ,d_\nu}-\frac{1}{2}\left( \delta^{2,2}\delta^{a_\nu ,d_\nu}\delta^{1,1}\delta^{b_\nu ,c_\nu}\right)\right\}\nonumber\\
&\hspace{30pt}+\frac{1}{2}\delta_{1,i'^2_\nu}\delta_{1,j'^2_\nu}\delta_{0,i'^1_\nu}\delta_{0,j'^1_\nu}\left(\delta^{a_\nu ,c_\nu}\right)+\frac{1}{2}\delta_{0,i'^2_\nu}\delta_{0,j'^2_\nu}\delta_{1,i'^1_\nu}\delta_{1,j'^1_\nu}\left(\delta^{b_\nu,d_\nu}\right)\nonumber\\
&\hspace{30pt}+\delta_{1,i'^2_\nu}\delta_{1,i'^1_\nu}\delta_{0,j'^2_\nu}\delta_{0,j'^1_\nu}\frac{1}{2!}\left(\epsilon^{21}\epsilon^{a_\nu b_\nu}\right)+\delta_{0,i'^2_\nu}\delta_{0,i'^1_\nu}\delta_{1,j'^2_\nu}\delta_{1,j'^1_\nu}\frac{1}{2!}\left(\epsilon^{c_\nu d_\nu}\epsilon^{21}\right)\bigg]\nonumber\\
&\hspace{30pt}\times g(i'^2_\nu)g(i'^1_\nu)g(j'^2_\nu)g(j'^1_\nu).
\end{align}
The difficulty in the Grassmann integration in Eq.~\eqref{eq:local_tensor2} is that we cannot simply align staggered fields $\chi^1,\chi^2,\bar{\chi}^1,\bar{\chi}^2$ to a set order. To see this, we rewrite Eq.~\eqref{eq:local_tensor2} in the following:
\begin{align}
        \mathcal{T}&=\sum_{\{i,j\}}\int  d\chi^1 d\bar{\chi}^1 d\chi^2 d\bar{\chi}^2 \left(1-m\bar{\chi}^1\chi^1-m\bar{\chi}^2\chi^2+m^2\bar{\chi}^1\chi^1\bar{\chi}^2\chi^2\right)\left(\prod_\nu \sum_{a_\nu,b_\nu,c_\nu,d_\nu}\right)\left(\prod_\nu F^{a_\nu b_\nu c_\nu d_\nu}_{i'^2_\nu i'^1_\nu j'^2_\nu j'^1_\nu}\right)\nonumber\\
    &\hspace{30pt}\times \left[\prod_\nu\left(\bar{\zeta}^2_\nu\chi^{a_\nu}\right)^{i'^2_\nu}\right]\left[\prod_\nu\left(\bar{\zeta}^1_\nu\chi^{b_\nu}\right)^{i'^1_\nu}\right]\left[\prod_\nu\left(\xi^1_\nu\chi^1\right)^{j^1_\nu}\right]\left[\prod_\nu\left(\xi^2_\nu\chi^2\right)^{j^2_\nu}\right]\nonumber\\
    &\hspace{30pt}\times \left[\prod_\nu \left(\bar{\chi}^{c_\nu}\bar{\xi}^2_\nu\right)^{j'^2_\nu}\right]\left[\prod_\nu\left(\bar{\chi}^{d_\nu}\bar{\xi}^1_\nu\right)^{j'^1_\nu}\right]\left[\prod_\nu\left(\bar{\chi}^1\zeta^1_\nu\right)^{i^1_\nu}\right]\left[\prod_\nu\left(\bar{\chi}^2\zeta^2_\nu\right)^{i^2_\nu}\right].\label{eq:T_append}
\end{align}
Examining Eq.~\eqref{eq:T_append}, the ordering of the $\chi^1,\chi^2$ and $\bar{\chi}^1,\bar{\chi}^2$ depends on the choice of pairs of the primed variables $i'_\nu,j'_\nu$ and the color indices $a_\nu,b_\nu,c_\nu,d_\nu$. 
Therefore, when performing the integration in the order $d\chi^1\, d\bar{\chi}^1\, d\chi^2\, d\bar{\chi}^2$, 
a straightforward strategy is to absorb the corresponding sign factors into 
$F^{a_\nu b_\nu c_\nu d_\nu}_{i'^2_\nu i'^1_\nu j'^2_\nu j'^1_\nu}$. 
The final form is given by
\begin{align*}
\tilde{F}^{a_\nu b_\nu c_\nu d_\nu}_{i'^2_\nu i'^1_\nu j'^2_\nu j'^1_\nu}&=\bigg[ \delta_{0,i'^2_\nu}\delta_{0,i'^1_\nu}\delta_{0,j'^2_\nu} \delta_{0,j'^1_\nu}+\delta_{1,i'^2_\nu}\delta_{1,i'^1_\nu}\delta_{1,j'^2_\nu}\delta_{1,j'^1_\nu}\frac{1}{3}\left\{\delta^{2,2}\delta^{a_\nu ,c_\nu}\delta^{1,1}\delta^{b_\nu ,d_\nu}+\frac{1}{2}\left( \delta^{2,2}\delta^{a_\nu ,d_\nu}\delta^{1,1}\delta^{b_\nu ,c_\nu}\right)\right\}\nonumber\\
&\hspace{30pt}+\frac{1}{2}\delta_{1,i'^2_\nu}\delta_{1,j'^2_\nu}\delta_{0,i'^1_\nu}\delta_{0,j'^1_\nu}\left(\delta^{a_\nu ,c_\nu}\right)+\frac{1}{2}\delta_{0,i'^2_\nu}\delta_{0,j'^2_\nu}\delta_{1,i'^1_\nu}\delta_{1,j'^1_\nu}\left(\delta^{b_\nu,d_\nu}\right)\nonumber\\
&\hspace{30pt}+\delta_{1,i'^2_\nu}\delta_{1,i'^1_\nu}\delta_{0,j'^2_\nu}\delta_{0,j'^1_\nu}\frac{1}{2!}\left(\epsilon^{21}|\epsilon^{a_\nu b_\nu}|\right)+\delta_{0,i'^2_\nu}\delta_{0,i'^1_\nu}\delta_{1,j'^2_\nu}\delta_{1,j'^1_\nu}\frac{1}{2!}\left(|\epsilon^{c_\nu d_\nu}|\epsilon^{21}\right)\bigg]\nonumber\\
&\hspace{30pt}\times g(i'^2_\nu)g(i'^1_\nu)g(j'^2_\nu)g(j'^1_\nu).
\end{align*}
By incorporating $\tilde{F}$ instead of $F$ into $\mathcal{T}$, 
we can effectively account for the complicated sign factors that arise in the integration.

After performing the integration of link variables and staggered fields, the auxiliary Grassmann variables $\zeta,\xi$ appear in the order
\begin{align}
&\left[\prod_\nu\left(\bar{\zeta}^2_\nu \right)^{i'^2_\nu}\right]
 \left[\prod_\nu\left(\bar{\zeta}^1_\nu \right)^{i'^1_\nu}\right]
 \left[\prod_\nu\left(\xi^1_\nu \right)^{j^1_\nu}\right]
 \left[\prod_\nu\left(\xi^2_\nu \right)^{j^2_\nu}\right]\nn\\
&\times 
 \left[\prod_\nu\left(\bar{\xi}^2_\nu\right)^{j'^2_\nu}\right]
 \left[\prod_\nu\left(\bar{\xi}^1_\nu\right)^{j'^1_\nu}\right]
 \left[\prod_\nu\left(\zeta^1_\nu\right)^{i^1_\nu}\right]
 \left[\prod_\nu\left(\zeta^2_\nu\right)^{i^2_\nu}\right].
\end{align}
To reorder them into the form of Eq.~\eqref{eq:GrassmanT}, 
we must include sign factors $(-1)^R$, where
\begin{align*}&R_{\left(i^1_1,i^2_1,j^1_1,j^2_1\right)\left(i^1_2,i^2_2,j^1_2,j^2_2\right)\left(i^1_3,i^2_3,j^1_3,j^2_3\right)\left(i^1_4,i^2_4,j^1_4,j^2_4\right)\left(i'^1_1,i'^2_1,j'^1_1,j'^2_1\right)\left(i'^1_2,i'^2_2,j'^1_2,j'^2_2\right)\left(i'^1_3,i'^2_3,j'^1_3,j'^2_3\right)\left(i'^1_4,i'^2_4,j'^1_4,j'^2_4\right)}\\
    &=i^1_1\left(\sum_{\nu=1}^4 j'^1_\nu+\sum_{\nu=1}^4 j'^2_\nu+\sum_{\nu=1}^4 j^1_\nu+\sum_{\nu=1}^4 j^2_\nu+\sum_{\nu=1}^4 i'^1_\nu+\sum_{\nu=1}^4 i'^2_\nu\right)\\
    &+i^2_1\left(\sum_{\nu=2}^4 i^1_\nu+\sum_{\nu=1}^4 j'^1_\nu+\sum_{\nu=1}^4 j'^2_\nu+\sum_{\nu=1}^4 j^2_\nu+\sum_{\nu=1}^4 j^1_\nu+\sum_{\nu=1}^4 i'^1_\nu+\sum_{\nu=1}^4 i'^2_\nu\right)\\
    &+j^1_1\left(\sum_{\nu=1}^4 i'^1_\nu+\sum_{\nu=1}^4 i'^2_\nu\right)\\
    &+j^2_1\left(\sum_{\nu=2}^4 j^1_\nu+\sum_{\nu=1}^4 i'^1_\nu+\sum_{\nu=1}^4 i'^2_\nu\right)+i^1_2\left(\sum_{\nu=1}^4 j'^1_\nu+\sum_{\nu=1}^4 j'^2_\nu+\sum_{\nu=2}^4 j^2_\nu+\sum_{\nu=2}^4 j^1_\nu+\sum_{\nu=1}^4 i'^1_\nu+\sum_{\nu=1}^4 i'^2_\nu\right)\\
    &+i^2_2\left(\sum_{\nu=3}^4 i^1_\nu+\sum_{\nu=1}^4 j'^1_\nu+\sum_{\nu=1}^4 j'^2_\nu+\sum_{\nu=2}^4 j^2_\nu+\sum_{\nu=2}^4 j^1_\nu+\sum_{\nu=1}^4 i'^1_\nu+\sum_{\nu=1}^4 i'^2_\nu\right)\\
    &+j^1_2\left(\sum_{\nu=1}^4 i'^1_\nu+\sum_{\nu=1}^4 i'^2_\nu\right)+j^2_2\left(\sum_{\nu=3}^4 j^1_\nu+\sum_{\nu=1}^4 i'^1_\nu+\sum_{\nu=1}^4 i'^2_\nu\right)\\
    &+i^1_3\left(\sum_{\nu=1}^4 j'^1_\nu+\sum_{\nu=1}^4 j'^2_\nu+\sum_{\nu=3}^4 j^2_\nu+\sum_{\nu=3}^4 j^1_\nu+\sum_{\nu=1}^4 i'^1_\nu+\sum_{\nu=1}^4 i'^2_\nu\right)\\
    &+i^2_3\left(i^1_4+\sum_{\nu=1}^4 j'^1_\nu+\sum_{\nu=1}^4 j'^2_\nu+\sum_{\nu=3}^4 j^2_\nu+\sum_{\nu=3}^4 j^1_\nu+\sum_{\nu=1}^4 i'^1_\nu+\sum_{\nu=1}^4 i'^2_\nu\right)\\
    &+j^1_3\left(\sum_{\nu=1}^4 i'^1_\nu+\sum_{\nu=1}^4 i'^2_\nu\right)+j^2_3\left(j^1_4+\sum_{\nu=1}^4 i'^1_\nu+\sum_{\nu=1}^4 i'^2_\nu\right)\\
    &+i^1_4\left(\sum_{\nu=1}^4 j'^1_\nu+\sum_{\nu=1}^4 j'^2_\nu+j^2_4+j^1_4+\sum_{\nu=1}^4 i'^1_\nu+\sum_{\nu=1}^4 i'^2_\nu\right)\\
    &+i^2_4\left(\sum_{\nu=1}^4 j'^1_\nu+\sum_{\nu=1}^4 j'^2_\nu+j^2_4+j^1_4+\sum_{\nu=1}^4 i'^1_\nu+\sum_{\nu=1}^4 i'^2_\nu\right)\\
    &+j^1_4\left(\sum_{\nu=1}^4 i'^1_\nu+\sum_{\nu=1}^4 i'^2_\nu\right)+j^2_4\left(\sum_{\nu=1}^4 i'^1_\nu+\sum_{\nu=1}^4 i'^2_\nu\right)\\
    &+j'^2_4\left(\sum_{\nu=1}^3 j'^2_\nu+\sum_{\nu=1}^4 i'^1_\nu+\sum_{\nu=1}^4 i'^2_\nu\right)+j'^1_4\left(\sum_{\nu=1}^3 j'^1_\nu+\sum_{\nu=1}^3 j'^2_\nu+\sum_{\nu=1}^4 i'^1_\nu+\sum_{\nu=1}^4 i'^2_\nu\right)\\
    &+i'^2_4\left(\sum_{\nu=1}^3 i'^2_\nu\right)+i'^1_4\left(\sum_{\nu=1}^3 i'^1_\nu+\sum_{\nu=1}^3 i'^2_\nu\right)\\
    &+j'^2_3\left(\sum_{\nu=1}^2 j'^2_\nu+\sum_{\nu=1}^3 i'^1_\nu+\sum_{\nu=1}^3 i'^2_\nu\right)+j'^1_3\left(\sum_{\nu=1}^2 j'^1_\nu+\sum_{\nu=1}^2 j'^2_\nu+\sum_{\nu=1}^3 i'^1_\nu+\sum_{\nu=1}^3 i'^2_\nu\right)+i'^2_3\left(\sum_{\nu=1}^2 i'^2_\nu\right)\\
    &+i'^1_3\left(\sum_{\nu=1}^2 i'^1_\nu+\sum_{\nu=1}^2 i'^2_\nu\right)\\
    &+j'^2_2\left(j'^2_1+\sum_{\nu=1}^2 i'^1_\nu+\sum_{\nu=1}^2 i'^2_\nu\right)\\
    &+j'^1_2\left(j'^1_1+j'^2_1+\sum_{\nu=1}^2 i'^1_\nu+\sum_{\nu=1}^2 i'^2_\nu\right)+i'^2_2 i'^2_1+i'^1_2\left(i'^1_1+i'^2_1\right)\\
    &+j'^2_1\left(i'^1_1+i'^2_1\right)+j'^1_1\left(i'^1_1+i'^2_1\right).
\end{align*}

Finally, we briefly comment on the numerical strategy for constructing the initial tensor in Eq.~\eqref{eq:coef}. Since the size of $T$ is $16^8$, it is impractical to implement a na\"{\i}ve for-loop over all indices. However, careful inspection of the Kronecker delta constraints arising from the fermion and link variable integrations in Eq.~\eqref{eq:coef} shows that they restrict nonzero contributions to only a small subset of entries, yielding a highly sparse and low-rank structure. Therefore, we performed for-loops only over the indices allowed by the Kronecker delta constraints and constructed $T$ in the form of a sparse matrix. To generate the fundamental tensors required in the ATRG method, we applied truncated SVD to $T$ using Arpack~\cite{doi:10.1137/1.9780898719628}. Once the five-leg tensors required for the subsequent ATRG calculation are constructed, they are treated as dense matrices endowed with a graded $\mathbb{Z}_{2}$ symmetry originating from the Grassmann parity throughout the coarse-graining procedure.


\bibliography{bib/formulation,bib/algorithm,bib/discrete,bib/grassmann,bib/continuous,bib/gauge,bib/real_time,bib/review,bib/for_this_paper}

\begin{thebibliography}{53}%
\makeatletter
\providecommand \@ifxundefined [1]{%
 \@ifx{#1\undefined}
}%
\providecommand \@ifnum [1]{%
 \ifnum #1\expandafter \@firstoftwo
 \else \expandafter \@secondoftwo
 \fi
}%
\providecommand \@ifx [1]{%
 \ifx #1\expandafter \@firstoftwo
 \else \expandafter \@secondoftwo
 \fi
}%
\providecommand \natexlab [1]{#1}%
\providecommand \enquote  [1]{``#1''}%
\providecommand \bibnamefont  [1]{#1}%
\providecommand \bibfnamefont [1]{#1}%
\providecommand \citenamefont [1]{#1}%
\providecommand \href@noop [0]{\@secondoftwo}%
\providecommand \href [0]{\begingroup \@sanitize@url \@href}%
\providecommand \@href[1]{\@@startlink{#1}\@@href}%
\providecommand \@@href[1]{\endgroup#1\@@endlink}%
\providecommand \@sanitize@url [0]{\catcode `\\12\catcode `\$12\catcode `\&12\catcode `\#12\catcode `\^12\catcode `\_12\catcode `\%12\relax}%
\providecommand \@@startlink[1]{}%
\providecommand \@@endlink[0]{}%
\providecommand \url  [0]{\begingroup\@sanitize@url \@url }%
\providecommand \@url [1]{\endgroup\@href {#1}{\urlprefix }}%
\providecommand \urlprefix  [0]{URL }%
\providecommand \Eprint [0]{\href }%
\providecommand \doibase [0]{https://doi.org/}%
\providecommand \selectlanguage [0]{\@gobble}%
\providecommand \bibinfo  [0]{\@secondoftwo}%
\providecommand \bibfield  [0]{\@secondoftwo}%
\providecommand \translation [1]{[#1]}%
\providecommand \BibitemOpen [0]{}%
\providecommand \bibitemStop [0]{}%
\providecommand \bibitemNoStop [0]{.\EOS\space}%
\providecommand \EOS [0]{\spacefactor3000\relax}%
\providecommand \BibitemShut  [1]{\csname bibitem#1\endcsname}%
\let\auto@bib@innerbib\@empty
\bibitem [{\citenamefont {Dagotto}\ \emph {et~al.}(1986)\citenamefont {Dagotto}, \citenamefont {Karsch},\ and\ \citenamefont {Moreo}}]{Dagotto:1986gw}%
  \BibitemOpen
  \bibfield  {author} {\bibinfo {author} {\bibfnamefont {E.}~\bibnamefont {Dagotto}}, \bibinfo {author} {\bibfnamefont {F.}~\bibnamefont {Karsch}},\ and\ \bibinfo {author} {\bibfnamefont {A.}~\bibnamefont {Moreo}},\ }\bibfield  {title} {\bibinfo {title} {{The Strong Coupling Limit of SU(2) {QCD} at Finite Baryon Density}},\ }\href {https://doi.org/10.1016/0370-2693(86)90383-7} {\bibfield  {journal} {\bibinfo  {journal} {Phys. Lett. B}\ }\textbf {\bibinfo {volume} {169}},\ \bibinfo {pages} {421} (\bibinfo {year} {1986})}\BibitemShut {NoStop}%
\bibitem [{\citenamefont {Dagotto}\ \emph {et~al.}(1987)\citenamefont {Dagotto}, \citenamefont {Moreo},\ and\ \citenamefont {Wolff}}]{Dagotto:1986xt}%
  \BibitemOpen
  \bibfield  {author} {\bibinfo {author} {\bibfnamefont {E.}~\bibnamefont {Dagotto}}, \bibinfo {author} {\bibfnamefont {A.}~\bibnamefont {Moreo}},\ and\ \bibinfo {author} {\bibfnamefont {U.}~\bibnamefont {Wolff}},\ }\bibfield  {title} {\bibinfo {title} {{Lattice SU($N$) {QCD} at Finite Temperature and Density in the Strong Coupling Limit}},\ }\href {https://doi.org/10.1016/0370-2693(87)90315-7} {\bibfield  {journal} {\bibinfo  {journal} {Phys. Lett. B}\ }\textbf {\bibinfo {volume} {186}},\ \bibinfo {pages} {395} (\bibinfo {year} {1987})}\BibitemShut {NoStop}%
\bibitem [{\citenamefont {Klatke}\ and\ \citenamefont {Mutter}(1990)}]{Klatke:1989xy}%
  \BibitemOpen
  \bibfield  {author} {\bibinfo {author} {\bibfnamefont {J.~U.}\ \bibnamefont {Klatke}}\ and\ \bibinfo {author} {\bibfnamefont {K.~H.}\ \bibnamefont {Mutter}},\ }\bibfield  {title} {\bibinfo {title} {{Strong Coupling {QCD} With SU(2) Gauge Fields at Finite Baryon Number Density}},\ }\href {https://doi.org/10.1016/0550-3213(90)90337-D} {\bibfield  {journal} {\bibinfo  {journal} {Nucl. Phys. B}\ }\textbf {\bibinfo {volume} {342}},\ \bibinfo {pages} {764} (\bibinfo {year} {1990})}\BibitemShut {NoStop}%
\bibitem [{\citenamefont {Nishida}\ \emph {et~al.}(2004)\citenamefont {Nishida}, \citenamefont {Fukushima},\ and\ \citenamefont {Hatsuda}}]{Nishida:2003uj}%
  \BibitemOpen
  \bibfield  {author} {\bibinfo {author} {\bibfnamefont {Y.}~\bibnamefont {Nishida}}, \bibinfo {author} {\bibfnamefont {K.}~\bibnamefont {Fukushima}},\ and\ \bibinfo {author} {\bibfnamefont {T.}~\bibnamefont {Hatsuda}},\ }\bibfield  {title} {\bibinfo {title} {{Thermodynamics of strong coupling two color QCD with chiral and diquark condensates}},\ }\href {https://doi.org/10.1016/j.physrep.2004.05.005} {\bibfield  {journal} {\bibinfo  {journal} {Phys. Rep.}\ }\textbf {\bibinfo {volume} {398}},\ \bibinfo {pages} {281} (\bibinfo {year} {2004})},\ \Eprint {https://arxiv.org/abs/hep-ph/0306066} {arXiv:hep-ph/0306066} \BibitemShut {NoStop}%
\bibitem [{\citenamefont {Nishida}(2004)}]{Nishida:2003fb}%
  \BibitemOpen
  \bibfield  {author} {\bibinfo {author} {\bibfnamefont {Y.}~\bibnamefont {Nishida}},\ }\bibfield  {title} {\bibinfo {title} {{Phase structures of strong coupling lattice QCD with finite baryon and isospin density}},\ }\href {https://doi.org/10.1103/PhysRevD.69.094501} {\bibfield  {journal} {\bibinfo  {journal} {Phys. Rev. D}\ }\textbf {\bibinfo {volume} {69}},\ \bibinfo {pages} {094501} (\bibinfo {year} {2004})},\ \Eprint {https://arxiv.org/abs/hep-ph/0312371} {arXiv:hep-ph/0312371} \BibitemShut {NoStop}%
\bibitem [{\citenamefont {Nakamura}(1984)}]{Nakamura:1984uz}%
  \BibitemOpen
  \bibfield  {author} {\bibinfo {author} {\bibfnamefont {A.}~\bibnamefont {Nakamura}},\ }\bibfield  {title} {\bibinfo {title} {{Quarks and Gluons at Finite Temperature and Density}},\ }\href {https://doi.org/10.1016/0370-2693(84)90430-1} {\bibfield  {journal} {\bibinfo  {journal} {Phys. Lett. B}\ }\textbf {\bibinfo {volume} {149}},\ \bibinfo {pages} {391} (\bibinfo {year} {1984})}\BibitemShut {NoStop}%
\bibitem [{\citenamefont {Boz}\ \emph {et~al.}(2020)\citenamefont {Boz}, \citenamefont {Giudice}, \citenamefont {Hands},\ and\ \citenamefont {Skullerud}}]{Boz:2019enj}%
  \BibitemOpen
  \bibfield  {author} {\bibinfo {author} {\bibfnamefont {T.}~\bibnamefont {Boz}}, \bibinfo {author} {\bibfnamefont {P.}~\bibnamefont {Giudice}}, \bibinfo {author} {\bibfnamefont {S.}~\bibnamefont {Hands}},\ and\ \bibinfo {author} {\bibfnamefont {J.-I.}\ \bibnamefont {Skullerud}},\ }\bibfield  {title} {\bibinfo {title} {{Dense two-color QCD towards continuum and chiral limits}},\ }\href {https://doi.org/10.1103/PhysRevD.101.074506} {\bibfield  {journal} {\bibinfo  {journal} {Phys. Rev. D}\ }\textbf {\bibinfo {volume} {101}},\ \bibinfo {pages} {074506} (\bibinfo {year} {2020})},\ \Eprint {https://arxiv.org/abs/1912.10975} {arXiv:1912.10975 [hep-lat]} \BibitemShut {NoStop}%
\bibitem [{\citenamefont {Begun}\ \emph {et~al.}(2022)\citenamefont {Begun}, \citenamefont {Bornyakov}, \citenamefont {Goy}, \citenamefont {Nakamura},\ and\ \citenamefont {Rogalyov}}]{Begun:2022bxj}%
  \BibitemOpen
  \bibfield  {author} {\bibinfo {author} {\bibfnamefont {A.}~\bibnamefont {Begun}}, \bibinfo {author} {\bibfnamefont {V.~G.}\ \bibnamefont {Bornyakov}}, \bibinfo {author} {\bibfnamefont {V.~A.}\ \bibnamefont {Goy}}, \bibinfo {author} {\bibfnamefont {A.}~\bibnamefont {Nakamura}},\ and\ \bibinfo {author} {\bibfnamefont {R.~N.}\ \bibnamefont {Rogalyov}},\ }\bibfield  {title} {\bibinfo {title} {{Study of two color QCD on large lattices}},\ }\href {https://doi.org/10.1103/PhysRevD.105.114505} {\bibfield  {journal} {\bibinfo  {journal} {Phys. Rev. D}\ }\textbf {\bibinfo {volume} {105}},\ \bibinfo {pages} {114505} (\bibinfo {year} {2022})},\ \Eprint {https://arxiv.org/abs/2203.04909} {arXiv:2203.04909 [hep-lat]} \BibitemShut {NoStop}%
\bibitem [{\citenamefont {Iida}\ \emph {et~al.}(2024)\citenamefont {Iida}, \citenamefont {Itou}, \citenamefont {Murakami},\ and\ \citenamefont {Suenaga}}]{Iida:2024irv}%
  \BibitemOpen
  \bibfield  {author} {\bibinfo {author} {\bibfnamefont {K.}~\bibnamefont {Iida}}, \bibinfo {author} {\bibfnamefont {E.}~\bibnamefont {Itou}}, \bibinfo {author} {\bibfnamefont {K.}~\bibnamefont {Murakami}},\ and\ \bibinfo {author} {\bibfnamefont {D.}~\bibnamefont {Suenaga}},\ }\bibfield  {title} {\bibinfo {title} {{Lattice study on finite density QC$_{2}$D towards zero temperature}},\ }\href {https://doi.org/10.1007/JHEP10(2024)022} {\bibfield  {journal} {\bibinfo  {journal} {JHEP}\ }\textbf {\bibinfo {volume} {10}},\ \bibinfo {pages} {022}},\ \Eprint {https://arxiv.org/abs/2405.20566} {arXiv:2405.20566 [hep-lat]} \BibitemShut {NoStop}%
\bibitem [{Note1()}]{Note1}%
  \BibitemOpen
  \bibinfo {note} {In this paper, the ``TRG method'' or the ``TRG approach'' refers to not only the original numerical algorithm proposed by Levin and Nave \cite {Levin:2006jai} but also its extensions~\cite {PhysRevB.86.045139,Shimizu:2014uva,PhysRevLett.115.180405,Sakai:2017jwp,PhysRevLett.118.110504,Hauru:2017tne,Adachi:2019paf,Kadoh:2019kqk,Akiyama:2020soe,PhysRevB.105.L060402,Akiyama:2022pse}.}\BibitemShut {Stop}%
\bibitem [{\citenamefont {Shimizu}\ and\ \citenamefont {Kuramashi}(2014{\natexlab{a}})}]{Shimizu:2014uva}%
  \BibitemOpen
  \bibfield  {author} {\bibinfo {author} {\bibfnamefont {Y.}~\bibnamefont {Shimizu}}\ and\ \bibinfo {author} {\bibfnamefont {Y.}~\bibnamefont {Kuramashi}},\ }\bibfield  {title} {\bibinfo {title} {{Grassmann tensor renormalization group approach to one-flavor lattice Schwinger model}},\ }\href {https://doi.org/10.1103/PhysRevD.90.014508} {\bibfield  {journal} {\bibinfo  {journal} {Phys. Rev.}\ }\textbf {\bibinfo {volume} {D90}},\ \bibinfo {pages} {014508} (\bibinfo {year} {2014}{\natexlab{a}})},\ \Eprint {https://arxiv.org/abs/1403.0642} {arXiv:1403.0642 [hep-lat]} \BibitemShut {NoStop}%
\bibitem [{\citenamefont {Shimizu}\ and\ \citenamefont {Kuramashi}(2014{\natexlab{b}})}]{Shimizu:2014fsa}%
  \BibitemOpen
  \bibfield  {author} {\bibinfo {author} {\bibfnamefont {Y.}~\bibnamefont {Shimizu}}\ and\ \bibinfo {author} {\bibfnamefont {Y.}~\bibnamefont {Kuramashi}},\ }\bibfield  {title} {\bibinfo {title} {{Critical behavior of the lattice Schwinger model with a topological term at $\theta=\pi$ using the Grassmann tensor renormalization group}},\ }\href {https://doi.org/10.1103/PhysRevD.90.074503} {\bibfield  {journal} {\bibinfo  {journal} {Phys. Rev.}\ }\textbf {\bibinfo {volume} {D90}},\ \bibinfo {pages} {074503} (\bibinfo {year} {2014}{\natexlab{b}})},\ \Eprint {https://arxiv.org/abs/1408.0897} {arXiv:1408.0897 [hep-lat]} \BibitemShut {NoStop}%
\bibitem [{\citenamefont {Kawauchi}\ and\ \citenamefont {Takeda}(2016)}]{Kawauchi:2016xng}%
  \BibitemOpen
  \bibfield  {author} {\bibinfo {author} {\bibfnamefont {H.}~\bibnamefont {Kawauchi}}\ and\ \bibinfo {author} {\bibfnamefont {S.}~\bibnamefont {Takeda}},\ }\bibfield  {title} {\bibinfo {title} {{Tensor renormalization group analysis of CP($N$-1) model}},\ }\href {https://doi.org/10.1103/PhysRevD.93.114503} {\bibfield  {journal} {\bibinfo  {journal} {Phys. Rev.}\ }\textbf {\bibinfo {volume} {D93}},\ \bibinfo {pages} {114503} (\bibinfo {year} {2016})},\ \Eprint {https://arxiv.org/abs/1603.09455} {arXiv:1603.09455 [hep-lat]} \BibitemShut {NoStop}%
\bibitem [{\citenamefont {Kawauchi}\ and\ \citenamefont {Takeda}(2018)}]{Kawauchi:2017dnj}%
  \BibitemOpen
  \bibfield  {author} {\bibinfo {author} {\bibfnamefont {H.}~\bibnamefont {Kawauchi}}\ and\ \bibinfo {author} {\bibfnamefont {S.}~\bibnamefont {Takeda}},\ }\bibfield  {title} {\bibinfo {title} {{Loop-TNR analysis of CP(1) model with theta term}},\ }\bibfield  {booktitle} {\emph {\bibinfo {booktitle} {{Proceedings, 35th International Symposium on Lattice Field Theory (Lattice 2017): Granada, Spain, June 18-24, 2017}}},\ }\href {https://doi.org/10.1051/epjconf/201817511015} {\bibfield  {journal} {\bibinfo  {journal} {EPJ Web Conf.}\ }\textbf {\bibinfo {volume} {175}},\ \bibinfo {pages} {11015} (\bibinfo {year} {2018})},\ \Eprint {https://arxiv.org/abs/1710.09804} {arXiv:1710.09804 [hep-lat]} \BibitemShut {NoStop}%
\bibitem [{\citenamefont {Yang}\ \emph {et~al.}(2016)\citenamefont {Yang}, \citenamefont {Liu}, \citenamefont {Zou}, \citenamefont {Xie},\ and\ \citenamefont {Meurice}}]{Yang:2015rra}%
  \BibitemOpen
  \bibfield  {author} {\bibinfo {author} {\bibfnamefont {L.-P.}\ \bibnamefont {Yang}}, \bibinfo {author} {\bibfnamefont {Y.}~\bibnamefont {Liu}}, \bibinfo {author} {\bibfnamefont {H.}~\bibnamefont {Zou}}, \bibinfo {author} {\bibfnamefont {Z.}~\bibnamefont {Xie}},\ and\ \bibinfo {author} {\bibfnamefont {Y.}~\bibnamefont {Meurice}},\ }\bibfield  {title} {\bibinfo {title} {{Fine structure of the entanglement entropy in the O(2) model}},\ }\href {https://doi.org/10.1103/PhysRevE.93.012138} {\bibfield  {journal} {\bibinfo  {journal} {Phys. Rev. E}\ }\textbf {\bibinfo {volume} {93}},\ \bibinfo {pages} {012138} (\bibinfo {year} {2016})},\ \Eprint {https://arxiv.org/abs/1507.01471} {arXiv:1507.01471 [cond-mat.stat-mech]} \BibitemShut {NoStop}%
\bibitem [{\citenamefont {Shimizu}\ and\ \citenamefont {Kuramashi}(2018)}]{Shimizu:2017onf}%
  \BibitemOpen
  \bibfield  {author} {\bibinfo {author} {\bibfnamefont {Y.}~\bibnamefont {Shimizu}}\ and\ \bibinfo {author} {\bibfnamefont {Y.}~\bibnamefont {Kuramashi}},\ }\bibfield  {title} {\bibinfo {title} {{Berezinskii-Kosterlitz-Thouless transition in lattice Schwinger model with one flavor of Wilson fermion}},\ }\href {https://doi.org/10.1103/PhysRevD.97.034502} {\bibfield  {journal} {\bibinfo  {journal} {Phys. Rev.}\ }\textbf {\bibinfo {volume} {D97}},\ \bibinfo {pages} {034502} (\bibinfo {year} {2018})},\ \Eprint {https://arxiv.org/abs/1712.07808} {arXiv:1712.07808 [hep-lat]} \BibitemShut {NoStop}%
\bibitem [{\citenamefont {Takeda}\ and\ \citenamefont {Yoshimura}(2015)}]{Takeda:2014vwa}%
  \BibitemOpen
  \bibfield  {author} {\bibinfo {author} {\bibfnamefont {S.}~\bibnamefont {Takeda}}\ and\ \bibinfo {author} {\bibfnamefont {Y.}~\bibnamefont {Yoshimura}},\ }\bibfield  {title} {\bibinfo {title} {{Grassmann tensor renormalization group for the one-flavor lattice Gross-Neveu model with finite chemical potential}},\ }\href {https://doi.org/10.1093/ptep/ptv022} {\bibfield  {journal} {\bibinfo  {journal} {PTEP}\ }\textbf {\bibinfo {volume} {2015}},\ \bibinfo {pages} {043B01} (\bibinfo {year} {2015})},\ \Eprint {https://arxiv.org/abs/1412.7855} {arXiv:1412.7855 [hep-lat]} \BibitemShut {NoStop}%
\bibitem [{\citenamefont {Kadoh}\ \emph {et~al.}(2018)\citenamefont {Kadoh}, \citenamefont {Kuramashi}, \citenamefont {Nakamura}, \citenamefont {Sakai}, \citenamefont {Takeda},\ and\ \citenamefont {Yoshimura}}]{Kadoh:2018hqq}%
  \BibitemOpen
  \bibfield  {author} {\bibinfo {author} {\bibfnamefont {D.}~\bibnamefont {Kadoh}}, \bibinfo {author} {\bibfnamefont {Y.}~\bibnamefont {Kuramashi}}, \bibinfo {author} {\bibfnamefont {Y.}~\bibnamefont {Nakamura}}, \bibinfo {author} {\bibfnamefont {R.}~\bibnamefont {Sakai}}, \bibinfo {author} {\bibfnamefont {S.}~\bibnamefont {Takeda}},\ and\ \bibinfo {author} {\bibfnamefont {Y.}~\bibnamefont {Yoshimura}},\ }\bibfield  {title} {\bibinfo {title} {{Tensor network formulation for two-dimensional lattice $ \mathcal{N} $ = 1 Wess-Zumino model}},\ }\href {https://doi.org/10.1007/JHEP03(2018)141} {\bibfield  {journal} {\bibinfo  {journal} {JHEP}\ }\textbf {\bibinfo {volume} {03}},\ \bibinfo {pages} {141}},\ \Eprint {https://arxiv.org/abs/1801.04183} {arXiv:1801.04183 [hep-lat]} \BibitemShut {NoStop}%
\bibitem [{\citenamefont {Kadoh}\ \emph {et~al.}(2020)\citenamefont {Kadoh}, \citenamefont {Kuramashi}, \citenamefont {Nakamura}, \citenamefont {Sakai}, \citenamefont {Takeda},\ and\ \citenamefont {Yoshimura}}]{Kadoh:2019ube}%
  \BibitemOpen
  \bibfield  {author} {\bibinfo {author} {\bibfnamefont {D.}~\bibnamefont {Kadoh}}, \bibinfo {author} {\bibfnamefont {Y.}~\bibnamefont {Kuramashi}}, \bibinfo {author} {\bibfnamefont {Y.}~\bibnamefont {Nakamura}}, \bibinfo {author} {\bibfnamefont {R.}~\bibnamefont {Sakai}}, \bibinfo {author} {\bibfnamefont {S.}~\bibnamefont {Takeda}},\ and\ \bibinfo {author} {\bibfnamefont {Y.}~\bibnamefont {Yoshimura}},\ }\bibfield  {title} {\bibinfo {title} {{Investigation of complex $\phi^{4}$ theory at finite density in two dimensions using TRG}},\ }\href {https://doi.org/10.1007/JHEP02(2020)161} {\bibfield  {journal} {\bibinfo  {journal} {JHEP}\ }\textbf {\bibinfo {volume} {02}},\ \bibinfo {pages} {161}},\ \Eprint {https://arxiv.org/abs/1912.13092} {arXiv:1912.13092 [hep-lat]} \BibitemShut {NoStop}%
\bibitem [{\citenamefont {Takeda}(2019)}]{Takeda:2019idb}%
  \BibitemOpen
  \bibfield  {author} {\bibinfo {author} {\bibfnamefont {S.}~\bibnamefont {Takeda}},\ }\bibfield  {title} {\bibinfo {title} {{Tensor network approach to real-time path integral}},\ }\href {https://doi.org/10.22323/1.363.0033} {\bibfield  {journal} {\bibinfo  {journal} {PoS}\ }\textbf {\bibinfo {volume} {LATTICE2019}},\ \bibinfo {pages} {033} (\bibinfo {year} {2019})},\ \Eprint {https://arxiv.org/abs/1908.00126} {arXiv:1908.00126 [hep-lat]} \BibitemShut {NoStop}%
\bibitem [{\citenamefont {Kuramashi}\ and\ \citenamefont {Yoshimura}(2020)}]{Kuramashi:2019cgs}%
  \BibitemOpen
  \bibfield  {author} {\bibinfo {author} {\bibfnamefont {Y.}~\bibnamefont {Kuramashi}}\ and\ \bibinfo {author} {\bibfnamefont {Y.}~\bibnamefont {Yoshimura}},\ }\bibfield  {title} {\bibinfo {title} {{Tensor renormalization group study of two-dimensional U(1) lattice gauge theory with a $\theta$ term}},\ }\href {https://doi.org/10.1007/JHEP04(2020)089} {\bibfield  {journal} {\bibinfo  {journal} {JHEP}\ }\textbf {\bibinfo {volume} {04}},\ \bibinfo {pages} {089}},\ \Eprint {https://arxiv.org/abs/1911.06480} {arXiv:1911.06480 [hep-lat]} \BibitemShut {NoStop}%
\bibitem [{\citenamefont {Akiyama}\ \emph {et~al.}(2020)\citenamefont {Akiyama}, \citenamefont {Kadoh}, \citenamefont {Kuramashi}, \citenamefont {Yamashita},\ and\ \citenamefont {Yoshimura}}]{Akiyama:2020ntf}%
  \BibitemOpen
  \bibfield  {author} {\bibinfo {author} {\bibfnamefont {S.}~\bibnamefont {Akiyama}}, \bibinfo {author} {\bibfnamefont {D.}~\bibnamefont {Kadoh}}, \bibinfo {author} {\bibfnamefont {Y.}~\bibnamefont {Kuramashi}}, \bibinfo {author} {\bibfnamefont {T.}~\bibnamefont {Yamashita}},\ and\ \bibinfo {author} {\bibfnamefont {Y.}~\bibnamefont {Yoshimura}},\ }\bibfield  {title} {\bibinfo {title} {{Tensor renormalization group approach to four-dimensional complex $\phi^4$ theory at finite density}},\ }\href {https://doi.org/10.1007/JHEP09(2020)177} {\bibfield  {journal} {\bibinfo  {journal} {JHEP}\ }\textbf {\bibinfo {volume} {09}},\ \bibinfo {pages} {177}},\ \Eprint {https://arxiv.org/abs/2005.04645} {arXiv:2005.04645 [hep-lat]} \BibitemShut {NoStop}%
\bibitem [{\citenamefont {Akiyama}\ \emph {et~al.}(2021)\citenamefont {Akiyama}, \citenamefont {Kuramashi}, \citenamefont {Yamashita},\ and\ \citenamefont {Yoshimura}}]{Akiyama:2020soe}%
  \BibitemOpen
  \bibfield  {author} {\bibinfo {author} {\bibfnamefont {S.}~\bibnamefont {Akiyama}}, \bibinfo {author} {\bibfnamefont {Y.}~\bibnamefont {Kuramashi}}, \bibinfo {author} {\bibfnamefont {T.}~\bibnamefont {Yamashita}},\ and\ \bibinfo {author} {\bibfnamefont {Y.}~\bibnamefont {Yoshimura}},\ }\bibfield  {title} {\bibinfo {title} {{Restoration of chiral symmetry in cold and dense Nambu--Jona-Lasinio model with tensor renormalization group}},\ }\href {https://doi.org/10.1007/JHEP01(2021)121} {\bibfield  {journal} {\bibinfo  {journal} {JHEP}\ }\textbf {\bibinfo {volume} {01}},\ \bibinfo {pages} {121}},\ \Eprint {https://arxiv.org/abs/2009.11583} {arXiv:2009.11583 [hep-lat]} \BibitemShut {NoStop}%
\bibitem [{\citenamefont {Nakayama}\ \emph {et~al.}(2022)\citenamefont {Nakayama}, \citenamefont {Funcke}, \citenamefont {Jansen}, \citenamefont {Kao},\ and\ \citenamefont {K\"uhn}}]{Nakayama:2021iyp}%
  \BibitemOpen
  \bibfield  {author} {\bibinfo {author} {\bibfnamefont {K.}~\bibnamefont {Nakayama}}, \bibinfo {author} {\bibfnamefont {L.}~\bibnamefont {Funcke}}, \bibinfo {author} {\bibfnamefont {K.}~\bibnamefont {Jansen}}, \bibinfo {author} {\bibfnamefont {Y.-J.}\ \bibnamefont {Kao}},\ and\ \bibinfo {author} {\bibfnamefont {S.}~\bibnamefont {K\"uhn}},\ }\bibfield  {title} {\bibinfo {title} {{Phase structure of the CP(1) model in the presence of a topological \ensuremath{\theta}-term}},\ }\href {https://doi.org/10.1103/PhysRevD.105.054507} {\bibfield  {journal} {\bibinfo  {journal} {Phys. Rev. D}\ }\textbf {\bibinfo {volume} {105}},\ \bibinfo {pages} {054507} (\bibinfo {year} {2022})},\ \Eprint {https://arxiv.org/abs/2107.14220} {arXiv:2107.14220 [hep-lat]} \BibitemShut {NoStop}%
\bibitem [{\citenamefont {Bloch}\ \emph {et~al.}(2021)\citenamefont {Bloch}, \citenamefont {Jha}, \citenamefont {Lohmayer},\ and\ \citenamefont {Meister}}]{Bloch:2021mjw}%
  \BibitemOpen
  \bibfield  {author} {\bibinfo {author} {\bibfnamefont {J.}~\bibnamefont {Bloch}}, \bibinfo {author} {\bibfnamefont {R.~G.}\ \bibnamefont {Jha}}, \bibinfo {author} {\bibfnamefont {R.}~\bibnamefont {Lohmayer}},\ and\ \bibinfo {author} {\bibfnamefont {M.}~\bibnamefont {Meister}},\ }\bibfield  {title} {\bibinfo {title} {{Tensor renormalization group study of the three-dimensional O(2) model}},\ }\href {https://doi.org/10.1103/PhysRevD.104.094517} {\bibfield  {journal} {\bibinfo  {journal} {Phys. Rev. D}\ }\textbf {\bibinfo {volume} {104}},\ \bibinfo {pages} {094517} (\bibinfo {year} {2021})},\ \Eprint {https://arxiv.org/abs/2105.08066} {arXiv:2105.08066 [hep-lat]} \BibitemShut {NoStop}%
\bibitem [{\citenamefont {Bloch}\ and\ \citenamefont {Lohmayer}(2023)}]{Bloch:2022vqz}%
  \BibitemOpen
  \bibfield  {author} {\bibinfo {author} {\bibfnamefont {J.}~\bibnamefont {Bloch}}\ and\ \bibinfo {author} {\bibfnamefont {R.}~\bibnamefont {Lohmayer}},\ }\bibfield  {title} {\bibinfo {title} {{Grassmann higher-order tensor renormalization group approach for two-dimensional strong-coupling QCD}},\ }\href {https://doi.org/10.1016/j.nuclphysb.2022.116032} {\bibfield  {journal} {\bibinfo  {journal} {Nucl. Phys. B}\ }\textbf {\bibinfo {volume} {986}},\ \bibinfo {pages} {116032} (\bibinfo {year} {2023})},\ \Eprint {https://arxiv.org/abs/2206.00545} {arXiv:2206.00545 [hep-lat]} \BibitemShut {NoStop}%
\bibitem [{\citenamefont {Akiyama}\ and\ \citenamefont {Kuramashi}(2023)}]{Akiyama:2023hvt}%
  \BibitemOpen
  \bibfield  {author} {\bibinfo {author} {\bibfnamefont {S.}~\bibnamefont {Akiyama}}\ and\ \bibinfo {author} {\bibfnamefont {Y.}~\bibnamefont {Kuramashi}},\ }\bibfield  {title} {\bibinfo {title} {{Critical endpoint of (3+1)-dimensional finite density \ensuremath{\mathbb{Z}}$_{3}$ gauge-Higgs model with tensor renormalization group}},\ }\href {https://doi.org/10.1007/JHEP10(2023)077} {\bibfield  {journal} {\bibinfo  {journal} {JHEP}\ }\textbf {\bibinfo {volume} {10}},\ \bibinfo {pages} {077}},\ \Eprint {https://arxiv.org/abs/2304.07934} {arXiv:2304.07934 [hep-lat]} \BibitemShut {NoStop}%
\bibitem [{\citenamefont {Akiyama}\ and\ \citenamefont {Kuramashi}(2024)}]{Akiyama:2024qer}%
  \BibitemOpen
  \bibfield  {author} {\bibinfo {author} {\bibfnamefont {S.}~\bibnamefont {Akiyama}}\ and\ \bibinfo {author} {\bibfnamefont {Y.}~\bibnamefont {Kuramashi}},\ }\bibfield  {title} {\bibinfo {title} {{Tensor renormalization group study of (1 + 1)-dimensional U(1) gauge-Higgs model at \ensuremath{\theta} = \ensuremath{\pi} with L\"uscher\textquoteright{}s admissibility condition}},\ }\href {https://doi.org/10.1007/JHEP09(2024)086} {\bibfield  {journal} {\bibinfo  {journal} {JHEP}\ }\textbf {\bibinfo {volume} {09}},\ \bibinfo {pages} {086}},\ \Eprint {https://arxiv.org/abs/2407.10409} {arXiv:2407.10409 [hep-lat]} \BibitemShut {NoStop}%
\bibitem [{\citenamefont {Luo}\ and\ \citenamefont {Kuramashi}(2023)}]{Luo:2022eje}%
  \BibitemOpen
  \bibfield  {author} {\bibinfo {author} {\bibfnamefont {X.}~\bibnamefont {Luo}}\ and\ \bibinfo {author} {\bibfnamefont {Y.}~\bibnamefont {Kuramashi}},\ }\bibfield  {title} {\bibinfo {title} {{Tensor renormalization group approach to (1+1)-dimensional SU(2) principal chiral model at finite density}},\ }\href {https://doi.org/10.1103/PhysRevD.107.094509} {\bibfield  {journal} {\bibinfo  {journal} {Phys. Rev. D}\ }\textbf {\bibinfo {volume} {107}},\ \bibinfo {pages} {094509} (\bibinfo {year} {2023})},\ \Eprint {https://arxiv.org/abs/2208.13991} {arXiv:2208.13991 [hep-lat]} \BibitemShut {NoStop}%
\bibitem [{\citenamefont {Hite}\ and\ \citenamefont {Meurice}(2025)}]{Hite:2024ulb}%
  \BibitemOpen
  \bibfield  {author} {\bibinfo {author} {\bibfnamefont {M.}~\bibnamefont {Hite}}\ and\ \bibinfo {author} {\bibfnamefont {Y.}~\bibnamefont {Meurice}},\ }\bibfield  {title} {\bibinfo {title} {{Quantum real-time evolution using tensor renormalization group methods}},\ }\href {https://doi.org/10.1103/PhysRevD.111.034502} {\bibfield  {journal} {\bibinfo  {journal} {Phys. Rev. D}\ }\textbf {\bibinfo {volume} {111}},\ \bibinfo {pages} {034502} (\bibinfo {year} {2025})},\ \Eprint {https://arxiv.org/abs/2411.05301} {arXiv:2411.05301 [quant-ph]} \BibitemShut {NoStop}%
\bibitem [{\citenamefont {Luo}\ and\ \citenamefont {Kuramashi}(2024)}]{Luo:2024lbh}%
  \BibitemOpen
  \bibfield  {author} {\bibinfo {author} {\bibfnamefont {X.}~\bibnamefont {Luo}}\ and\ \bibinfo {author} {\bibfnamefont {Y.}~\bibnamefont {Kuramashi}},\ }\bibfield  {title} {\bibinfo {title} {{Quantum phase transition of (1+1)-dimensional O(3) nonlinear sigma model at finite density with tensor renormalization group}},\ }\href {https://doi.org/10.1007/JHEP11(2024)144} {\bibfield  {journal} {\bibinfo  {journal} {JHEP}\ }\textbf {\bibinfo {volume} {11}},\ \bibinfo {pages} {144}},\ \Eprint {https://arxiv.org/abs/2406.08865} {arXiv:2406.08865 [hep-lat]} \BibitemShut {NoStop}%
\bibitem [{\citenamefont {Luo}\ and\ \citenamefont {Kuramashi}(2025)}]{Luo:2025qtv}%
  \BibitemOpen
  \bibfield  {author} {\bibinfo {author} {\bibfnamefont {X.}~\bibnamefont {Luo}}\ and\ \bibinfo {author} {\bibfnamefont {Y.}~\bibnamefont {Kuramashi}},\ }\bibfield  {title} {\bibinfo {title} {{Critical endpoints of three-dimensional finite density SU(3) spin model with tensor renormalization group}},\ }\href {https://doi.org/10.1007/JHEP07(2025)036} {\bibfield  {journal} {\bibinfo  {journal} {JHEP}\ }\textbf {\bibinfo {volume} {07}},\ \bibinfo {pages} {036}},\ \Eprint {https://arxiv.org/abs/2503.05144} {arXiv:2503.05144 [hep-lat]} \BibitemShut {NoStop}%
\bibitem [{\citenamefont {Asaduzzaman}\ \emph {et~al.}(2024)\citenamefont {Asaduzzaman}, \citenamefont {Catterall}, \citenamefont {Meurice}, \citenamefont {Sakai},\ and\ \citenamefont {Toga}}]{Asaduzzaman:2023pyz}%
  \BibitemOpen
  \bibfield  {author} {\bibinfo {author} {\bibfnamefont {M.}~\bibnamefont {Asaduzzaman}}, \bibinfo {author} {\bibfnamefont {S.}~\bibnamefont {Catterall}}, \bibinfo {author} {\bibfnamefont {Y.}~\bibnamefont {Meurice}}, \bibinfo {author} {\bibfnamefont {R.}~\bibnamefont {Sakai}},\ and\ \bibinfo {author} {\bibfnamefont {G.~C.}\ \bibnamefont {Toga}},\ }\bibfield  {title} {\bibinfo {title} {{Tensor network representation of non-abelian gauge theory coupled to reduced staggered fermions}},\ }\href {https://doi.org/10.1007/JHEP05(2024)195} {\bibfield  {journal} {\bibinfo  {journal} {JHEP}\ }\textbf {\bibinfo {volume} {05}},\ \bibinfo {pages} {195}},\ \Eprint {https://arxiv.org/abs/2312.16167} {arXiv:2312.16167 [hep-lat]} \BibitemShut {NoStop}%
\bibitem [{\citenamefont {Pai}\ \emph {et~al.}(2025)\citenamefont {Pai}, \citenamefont {Akiyama},\ and\ \citenamefont {Todo}}]{Pai:2024tip}%
  \BibitemOpen
  \bibfield  {author} {\bibinfo {author} {\bibfnamefont {K.~H.}\ \bibnamefont {Pai}}, \bibinfo {author} {\bibfnamefont {S.}~\bibnamefont {Akiyama}},\ and\ \bibinfo {author} {\bibfnamefont {S.}~\bibnamefont {Todo}},\ }\bibfield  {title} {\bibinfo {title} {{Grassmann tensor renormalization group approach to (1+1)-dimensional two-color lattice QCD at finite density}},\ }\href {https://doi.org/10.1007/JHEP03(2025)027} {\bibfield  {journal} {\bibinfo  {journal} {JHEP}\ }\textbf {\bibinfo {volume} {03}},\ \bibinfo {pages} {027}},\ \Eprint {https://arxiv.org/abs/2410.09485} {arXiv:2410.09485 [hep-lat]} \BibitemShut {NoStop}%
\bibitem [{\citenamefont {Adachi}\ \emph {et~al.}(2022)\citenamefont {Adachi}, \citenamefont {Okubo},\ and\ \citenamefont {Todo}}]{PhysRevB.105.L060402}%
  \BibitemOpen
  \bibfield  {author} {\bibinfo {author} {\bibfnamefont {D.}~\bibnamefont {Adachi}}, \bibinfo {author} {\bibfnamefont {T.}~\bibnamefont {Okubo}},\ and\ \bibinfo {author} {\bibfnamefont {S.}~\bibnamefont {Todo}},\ }\bibfield  {title} {\bibinfo {title} {Bond-weighted tensor renormalization group},\ }\href {https://doi.org/10.1103/PhysRevB.105.L060402} {\bibfield  {journal} {\bibinfo  {journal} {Phys. Rev. B}\ }\textbf {\bibinfo {volume} {105}},\ \bibinfo {pages} {L060402} (\bibinfo {year} {2022})},\ \Eprint {https://arxiv.org/abs/2011.01679} {arXiv:2011.01679 [cond-mat.stat-mech]} \BibitemShut {NoStop}%
\bibitem [{\citenamefont {Akiyama}(2022)}]{Akiyama:2022pse}%
  \BibitemOpen
  \bibfield  {author} {\bibinfo {author} {\bibfnamefont {S.}~\bibnamefont {Akiyama}},\ }\bibfield  {title} {\bibinfo {title} {{Bond-weighting method for the Grassmann tensor renormalization group}},\ }\href {https://doi.org/10.1007/JHEP11(2022)030} {\bibfield  {journal} {\bibinfo  {journal} {JHEP}\ }\textbf {\bibinfo {volume} {11}},\ \bibinfo {pages} {030}},\ \Eprint {https://arxiv.org/abs/2208.03227} {arXiv:2208.03227 [hep-lat]} \BibitemShut {NoStop}%
\bibitem [{\citenamefont {Adachi}\ \emph {et~al.}(2020)\citenamefont {Adachi}, \citenamefont {Okubo},\ and\ \citenamefont {Todo}}]{Adachi:2019paf}%
  \BibitemOpen
  \bibfield  {author} {\bibinfo {author} {\bibfnamefont {D.}~\bibnamefont {Adachi}}, \bibinfo {author} {\bibfnamefont {T.}~\bibnamefont {Okubo}},\ and\ \bibinfo {author} {\bibfnamefont {S.}~\bibnamefont {Todo}},\ }\bibfield  {title} {\bibinfo {title} {{Anisotropic Tensor Renormalization Group}},\ }\href {https://doi.org/10.1103/PhysRevB.102.054432} {\bibfield  {journal} {\bibinfo  {journal} {Phys. Rev. B}\ }\textbf {\bibinfo {volume} {102}},\ \bibinfo {pages} {054432} (\bibinfo {year} {2020})},\ \Eprint {https://arxiv.org/abs/1906.02007} {arXiv:1906.02007 [cond-mat.stat-mech]} \BibitemShut {NoStop}%
\bibitem [{\citenamefont {Oba}(2020)}]{Oba:2019csk}%
  \BibitemOpen
  \bibfield  {author} {\bibinfo {author} {\bibfnamefont {H.}~\bibnamefont {Oba}},\ }\bibfield  {title} {\bibinfo {title} {{Cost Reduction of Swapping Bonds Part in Anisotropic Tensor Renormalization Group}},\ }\href {https://doi.org/10.1093/ptep/ptz133} {\bibfield  {journal} {\bibinfo  {journal} {PTEP}\ }\textbf {\bibinfo {volume} {2020}},\ \bibinfo {pages} {013B02} (\bibinfo {year} {2020})},\ \Eprint {https://arxiv.org/abs/1908.07295} {arXiv:1908.07295 [cond-mat.stat-mech]} \BibitemShut {NoStop}%
\bibitem [{\citenamefont {Hands}\ \emph {et~al.}(1999)\citenamefont {Hands}, \citenamefont {Kogut}, \citenamefont {Lombardo},\ and\ \citenamefont {Morrison}}]{Hands:1999md}%
  \BibitemOpen
  \bibfield  {author} {\bibinfo {author} {\bibfnamefont {S.}~\bibnamefont {Hands}}, \bibinfo {author} {\bibfnamefont {J.~B.}\ \bibnamefont {Kogut}}, \bibinfo {author} {\bibfnamefont {M.-P.}\ \bibnamefont {Lombardo}},\ and\ \bibinfo {author} {\bibfnamefont {S.~E.}\ \bibnamefont {Morrison}},\ }\bibfield  {title} {\bibinfo {title} {{Symmetries and spectrum of SU(2) lattice gauge theory at finite chemical potential}},\ }\href {https://doi.org/10.1016/S0550-3213(99)00364-8} {\bibfield  {journal} {\bibinfo  {journal} {Nucl. Phys. B}\ }\textbf {\bibinfo {volume} {558}},\ \bibinfo {pages} {327} (\bibinfo {year} {1999})},\ \Eprint {https://arxiv.org/abs/hep-lat/9902034} {arXiv:hep-lat/9902034} \BibitemShut {NoStop}%
\bibitem [{\citenamefont {Pauli}(1957)}]{Pauli:1957voo}%
  \BibitemOpen
  \bibfield  {author} {\bibinfo {author} {\bibfnamefont {W.}~\bibnamefont {Pauli}},\ }\bibfield  {title} {\bibinfo {title} {{On the conservation of the Lepton charge}},\ }\href {https://doi.org/10.1007/bf02827771} {\bibfield  {journal} {\bibinfo  {journal} {Nuovo Cim.}\ }\textbf {\bibinfo {volume} {6}},\ \bibinfo {pages} {204} (\bibinfo {year} {1957})}\BibitemShut {NoStop}%
\bibitem [{\citenamefont {G{\"u}rsey}(1958)}]{Gursey:1958fzy}%
  \BibitemOpen
  \bibfield  {author} {\bibinfo {author} {\bibfnamefont {F.}~\bibnamefont {G{\"u}rsey}},\ }\bibfield  {title} {\bibinfo {title} {{Relation of charge independence and baryon conservation to Pauli{\textquoteright}s transformation}},\ }\href {https://doi.org/10.1007/bf02747705} {\bibfield  {journal} {\bibinfo  {journal} {Nuovo Cim.}\ }\textbf {\bibinfo {volume} {7}},\ \bibinfo {pages} {411} (\bibinfo {year} {1958})}\BibitemShut {NoStop}%
\bibitem [{\citenamefont {Kogut}\ \emph {et~al.}(2001)\citenamefont {Kogut}, \citenamefont {Sinclair}, \citenamefont {Hands},\ and\ \citenamefont {Morrison}}]{Kogut:2001na}%
  \BibitemOpen
  \bibfield  {author} {\bibinfo {author} {\bibfnamefont {J.~B.}\ \bibnamefont {Kogut}}, \bibinfo {author} {\bibfnamefont {D.~K.}\ \bibnamefont {Sinclair}}, \bibinfo {author} {\bibfnamefont {S.~J.}\ \bibnamefont {Hands}},\ and\ \bibinfo {author} {\bibfnamefont {S.~E.}\ \bibnamefont {Morrison}},\ }\bibfield  {title} {\bibinfo {title} {{Two color QCD at nonzero quark number density}},\ }\href {https://doi.org/10.1103/PhysRevD.64.094505} {\bibfield  {journal} {\bibinfo  {journal} {Phys. Rev. D}\ }\textbf {\bibinfo {volume} {64}},\ \bibinfo {pages} {094505} (\bibinfo {year} {2001})},\ \Eprint {https://arxiv.org/abs/hep-lat/0105026} {arXiv:hep-lat/0105026} \BibitemShut {NoStop}%
\bibitem [{\citenamefont {Akiyama}\ and\ \citenamefont {Kadoh}(2021)}]{Akiyama:2020sfo}%
  \BibitemOpen
  \bibfield  {author} {\bibinfo {author} {\bibfnamefont {S.}~\bibnamefont {Akiyama}}\ and\ \bibinfo {author} {\bibfnamefont {D.}~\bibnamefont {Kadoh}},\ }\bibfield  {title} {\bibinfo {title} {{More about the Grassmann tensor renormalization group}},\ }\href {https://doi.org/10.1007/JHEP10(2021)188} {\bibfield  {journal} {\bibinfo  {journal} {JHEP}\ }\textbf {\bibinfo {volume} {10}},\ \bibinfo {pages} {188}},\ \Eprint {https://arxiv.org/abs/2005.07570} {arXiv:2005.07570 [hep-lat]} \BibitemShut {NoStop}%
\bibitem [{\citenamefont {Weingarten}(1978)}]{Weingarten:1977ya}%
  \BibitemOpen
  \bibfield  {author} {\bibinfo {author} {\bibfnamefont {D.}~\bibnamefont {Weingarten}},\ }\bibfield  {title} {\bibinfo {title} {{Asymptotic Behavior of Group Integrals in the Limit of Infinite Rank}},\ }\href {https://doi.org/10.1063/1.523807} {\bibfield  {journal} {\bibinfo  {journal} {J. Math. Phys.}\ }\textbf {\bibinfo {volume} {19}},\ \bibinfo {pages} {999} (\bibinfo {year} {1978})}\BibitemShut {NoStop}%
\bibitem [{\citenamefont {Sugimoto}\ and\ \citenamefont {Sasaki}(2025)}]{Sugimoto:2025xva}%
  \BibitemOpen
  \bibfield  {author} {\bibinfo {author} {\bibfnamefont {Y.}~\bibnamefont {Sugimoto}}\ and\ \bibinfo {author} {\bibfnamefont {S.}~\bibnamefont {Sasaki}},\ }\bibfield  {title} {\bibinfo {title} {{Triad representation for the anisotropic tensor renormalization group in four dimensions}},\ }\href {https://doi.org/10.1103/c3qc-tn48} {\bibfield  {journal} {\bibinfo  {journal} {Phys. Rev. D}\ }\textbf {\bibinfo {volume} {112}},\ \bibinfo {pages} {094514} (\bibinfo {year} {2025})},\ \Eprint {https://arxiv.org/abs/2507.21909} {arXiv:2507.21909 [hep-lat]} \BibitemShut {NoStop}%
\bibitem [{\citenamefont {Lehoucq}\ \emph {et~al.}(1998)\citenamefont {Lehoucq}, \citenamefont {Sorensen},\ and\ \citenamefont {Yang}}]{doi:10.1137/1.9780898719628}%
  \BibitemOpen
  \bibfield  {author} {\bibinfo {author} {\bibfnamefont {R.~B.}\ \bibnamefont {Lehoucq}}, \bibinfo {author} {\bibfnamefont {D.~C.}\ \bibnamefont {Sorensen}},\ and\ \bibinfo {author} {\bibfnamefont {C.}~\bibnamefont {Yang}},\ }\href {https://doi.org/10.1137/1.9780898719628} {\emph {\bibinfo {title} {ARPACK Users' Guide}}}\ (\bibinfo  {publisher} {Society for Industrial and Applied Mathematics},\ \bibinfo {year} {1998})\BibitemShut {NoStop}%
\bibitem [{\citenamefont {Levin}\ and\ \citenamefont {Nave}(2007)}]{Levin:2006jai}%
  \BibitemOpen
  \bibfield  {author} {\bibinfo {author} {\bibfnamefont {M.}~\bibnamefont {Levin}}\ and\ \bibinfo {author} {\bibfnamefont {C.~P.}\ \bibnamefont {Nave}},\ }\bibfield  {title} {\bibinfo {title} {{Tensor renormalization group approach to two-dimensional classical lattice models}},\ }\href {https://doi.org/10.1103/PhysRevLett.99.120601} {\bibfield  {journal} {\bibinfo  {journal} {Phys. Rev. Lett.}\ }\textbf {\bibinfo {volume} {99}},\ \bibinfo {pages} {120601} (\bibinfo {year} {2007})},\ \Eprint {https://arxiv.org/abs/cond-mat/0611687} {arXiv:cond-mat/0611687 [cond-mat.stat-mech]} \BibitemShut {NoStop}%
\bibitem [{\citenamefont {Xie}\ \emph {et~al.}(2012)\citenamefont {Xie}, \citenamefont {Chen}, \citenamefont {Qin}, \citenamefont {Zhu}, \citenamefont {Yang},\ and\ \citenamefont {Xiang}}]{PhysRevB.86.045139}%
  \BibitemOpen
  \bibfield  {author} {\bibinfo {author} {\bibfnamefont {Z.~Y.}\ \bibnamefont {Xie}}, \bibinfo {author} {\bibfnamefont {J.}~\bibnamefont {Chen}}, \bibinfo {author} {\bibfnamefont {M.~P.}\ \bibnamefont {Qin}}, \bibinfo {author} {\bibfnamefont {J.~W.}\ \bibnamefont {Zhu}}, \bibinfo {author} {\bibfnamefont {L.~P.}\ \bibnamefont {Yang}},\ and\ \bibinfo {author} {\bibfnamefont {T.}~\bibnamefont {Xiang}},\ }\bibfield  {title} {\bibinfo {title} {Coarse-graining renormalization by higher-order singular value decomposition},\ }\href {https://doi.org/10.1103/PhysRevB.86.045139} {\bibfield  {journal} {\bibinfo  {journal} {Phys. Rev. B}\ }\textbf {\bibinfo {volume} {86}},\ \bibinfo {pages} {045139} (\bibinfo {year} {2012})},\ \Eprint {https://arxiv.org/abs/1201.1144} {arXiv:1201.1144 [cond-mat.stat-mech]} \BibitemShut {NoStop}%
\bibitem [{\citenamefont {Evenbly}\ and\ \citenamefont {Vidal}(2015)}]{PhysRevLett.115.180405}%
  \BibitemOpen
  \bibfield  {author} {\bibinfo {author} {\bibfnamefont {G.}~\bibnamefont {Evenbly}}\ and\ \bibinfo {author} {\bibfnamefont {G.}~\bibnamefont {Vidal}},\ }\bibfield  {title} {\bibinfo {title} {Tensor network renormalization},\ }\href {https://doi.org/10.1103/PhysRevLett.115.180405} {\bibfield  {journal} {\bibinfo  {journal} {Phys. Rev. Lett.}\ }\textbf {\bibinfo {volume} {115}},\ \bibinfo {pages} {180405} (\bibinfo {year} {2015})}\BibitemShut {NoStop}%
\bibitem [{\citenamefont {Sakai}\ \emph {et~al.}(2017)\citenamefont {Sakai}, \citenamefont {Takeda},\ and\ \citenamefont {Yoshimura}}]{Sakai:2017jwp}%
  \BibitemOpen
  \bibfield  {author} {\bibinfo {author} {\bibfnamefont {R.}~\bibnamefont {Sakai}}, \bibinfo {author} {\bibfnamefont {S.}~\bibnamefont {Takeda}},\ and\ \bibinfo {author} {\bibfnamefont {Y.}~\bibnamefont {Yoshimura}},\ }\bibfield  {title} {\bibinfo {title} {{Higher order tensor renormalization group for relativistic fermion systems}},\ }\href {https://doi.org/10.1093/ptep/ptx080} {\bibfield  {journal} {\bibinfo  {journal} {PTEP}\ }\textbf {\bibinfo {volume} {2017}},\ \bibinfo {pages} {063B07} (\bibinfo {year} {2017})},\ \Eprint {https://arxiv.org/abs/1705.07764} {arXiv:1705.07764 [hep-lat]} \BibitemShut {NoStop}%
\bibitem [{\citenamefont {Yang}\ \emph {et~al.}(2017)\citenamefont {Yang}, \citenamefont {Gu},\ and\ \citenamefont {Wen}}]{PhysRevLett.118.110504}%
  \BibitemOpen
  \bibfield  {author} {\bibinfo {author} {\bibfnamefont {S.}~\bibnamefont {Yang}}, \bibinfo {author} {\bibfnamefont {Z.-C.}\ \bibnamefont {Gu}},\ and\ \bibinfo {author} {\bibfnamefont {X.-G.}\ \bibnamefont {Wen}},\ }\bibfield  {title} {\bibinfo {title} {Loop optimization for tensor network renormalization},\ }\href {https://doi.org/10.1103/PhysRevLett.118.110504} {\bibfield  {journal} {\bibinfo  {journal} {Phys. Rev. Lett.}\ }\textbf {\bibinfo {volume} {118}},\ \bibinfo {pages} {110504} (\bibinfo {year} {2017})}\BibitemShut {NoStop}%
\bibitem [{\citenamefont {Hauru}\ \emph {et~al.}(2018)\citenamefont {Hauru}, \citenamefont {Delcamp},\ and\ \citenamefont {Mizera}}]{Hauru:2017tne}%
  \BibitemOpen
  \bibfield  {author} {\bibinfo {author} {\bibfnamefont {M.}~\bibnamefont {Hauru}}, \bibinfo {author} {\bibfnamefont {C.}~\bibnamefont {Delcamp}},\ and\ \bibinfo {author} {\bibfnamefont {S.}~\bibnamefont {Mizera}},\ }\bibfield  {title} {\bibinfo {title} {{Renormalization of tensor networks using graph independent local truncations}},\ }\href {https://doi.org/10.1103/PhysRevB.97.045111} {\bibfield  {journal} {\bibinfo  {journal} {Phys. Rev.}\ }\textbf {\bibinfo {volume} {B97}},\ \bibinfo {pages} {045111} (\bibinfo {year} {2018})},\ \Eprint {https://arxiv.org/abs/1709.07460} {arXiv:1709.07460 [cond-mat.str-el]} \BibitemShut {NoStop}%
\bibitem [{\citenamefont {Kadoh}\ and\ \citenamefont {Nakayama}(2019)}]{Kadoh:2019kqk}%
  \BibitemOpen
  \bibfield  {author} {\bibinfo {author} {\bibfnamefont {D.}~\bibnamefont {Kadoh}}\ and\ \bibinfo {author} {\bibfnamefont {K.}~\bibnamefont {Nakayama}},\ }\href@noop {} {\bibinfo {title} {{Renormalization group on a triad network}}} (\bibinfo {year} {2019}),\ \Eprint {https://arxiv.org/abs/1912.02414} {arXiv:1912.02414 [hep-lat]} \BibitemShut {NoStop}%
\end{thebibliography}%

\end{document}